\documentclass[12pt]{article}%
\usepackage[nosort]{cite}
\usepackage{graphicx}
\usepackage{multicol}
\usepackage{amsfonts}
\usepackage{amsmath}
\usepackage{amssymb}
\usepackage{amsthm}
\usepackage{heck}
\usepackage{afterpage}
\usepackage{setspace}
\usepackage{verbatim}
\usepackage{color}
\usepackage{longtable}
\usepackage{float}
\usepackage{subcaption}
\usepackage{epsfig}
\usepackage{epstopdf}
\usepackage{adjustbox}
\usepackage{tikz}
\usepackage[margin=1in]{geometry}
\usepackage{titletoc}
\usepackage{hyperref}%
\usepackage{mathrsfs}
\usepackage{dsfont}
\usepackage{algorithm}
\usepackage{algpseudocode}
\usepackage{pgfplots}
\usepackage{algorithmicx, algpseudocode,algorithm}
\usepackage{caption}
\usepackage{float}
\usepackage{makecell}
\usepackage{mathtools}
\usepackage{pdflscape}
\usepackage{array}
\usepackage{longtable}
\usepackage{comment}
\usepackage{mystuff}

\numberwithin{equation}{section}
\newsavebox{\mysavebox}

\hypersetup{colorlinks,citecolor=black,filecolor=black,linkcolor=black,urlcolor=black}

\usepackage[capitalize, noabbrev]{cleveref}
\usetikzlibrary{decorations.markings}
\usetikzlibrary{decorations.markings}
\usetikzlibrary{hobby}
\usepackage{tikz}
\usepackage{tikz-cd}
\usetikzlibrary{arrows}
\usetikzlibrary{arrows.meta}
\usetikzlibrary{arrows,decorations.pathmorphing}
\usetikzlibrary{shapes.geometric,calc,arrows, positioning,shapes.misc,decorations.markings}
\tikzset{
  big arrow/.style={
    decoration={markings,mark=at position 1 with {\arrow[scale=2,#1]{>}}},
    postaction={decorate},
    shorten >=0.4pt},
  big arrow/.default=black}

\pgfdeclarelayer{edgelayer}
\pgfdeclarelayer{nodelayer}
\pgfsetlayers{edgelayer,nodelayer,main}
\pgfplotsset{compat=1.16}
\tikzstyle{none}=[inner sep=0pt]

\theoremstyle{definition}

\crefname{thm}{Theorem}{Theorems}
\crefname{prop}{Proposition}{Propositions}
\crefname{defn}{Definition}{Definitions}
\crefname{lem}{Lemma}{Lemmas}

\tikzstyle{NodeCross}=[draw, shape=circle, cross out, inner sep=0pt, minimum size=6pt,line width=0.25mm]
\tikzstyle{Circle}=[draw, shape=circle, black, fill=black, inner sep=0pt, minimum size=6pt]
\tikzstyle{circle}=[draw, shape=circle, black, fill=black, inner sep=0pt, minimum size=16pt]
\tikzstyle{Star}=[draw, shape=star, fill=black, star points=8, inner sep=0pt, minimum size=8pt]
\tikzstyle{CircleRed}=[draw, shape=circle, black, fill=red, inner sep=0pt, minimum size=6pt]
\tikzstyle{StarP}=[draw={rgb,255: red,128; green,0; blue,128}, shape=star, fill={rgb,256: red,128; green,0; blue,128}, star points=8, inner sep=0pt, minimum size=12pt]
\tikzstyle{ShadedCircRed}=[draw=red, shape=circle, fill={rgb, 255: red,255; green,114; blue, 118}, inner sep=0pt, minimum size=80pt, line width=0.5mm, fill opacity=0.2]
\tikzstyle{ShadedCircRed2}=[draw=red, shape=circle, fill={rgb, 255: red,255; green,114; blue, 118}, inner sep=0pt, minimum size=10pt]
\tikzstyle{ShadedCircRed3}=[draw=black, shape=rectangle, fill={rgb, 255: red,255; green,114; blue, 118}, inner sep=0pt, minimum size=113pt, line width=0.25mm]
\tikzstyle{ShadedCirc}=[draw=red, shape=circle, fill=white, inner sep=0pt, minimum size=45pt,  fill opacity=1.0,  line width=0.5mm]
\tikzstyle{CircleBlue}=[draw, shape=circle, fill=blue, inner sep=0pt, minimum size=6pt]
\tikzstyle{BigCirclePurple}=[draw, shape=circle, fill={rgb,255: red,191; green,0; blue,191}, inner sep=0pt, minimum size=12pt]
\tikzstyle{CirclePurple}=[draw, shape=circle, fill={rgb,255: red,191; green,0; blue,191}, inner sep=0pt, minimum size=5pt]
\tikzstyle{EmptyCircle}=[draw, shape=circle, inner sep=0pt, minimum size=4pt]
\tikzstyle{GreenCircle}=[draw, shape=circle,  fill={rgb,255: red,80; green,200; blue,120}, inner sep=0pt, minimum size=8pt]
\tikzstyle{BrownCircle}=[draw, shape=circle,  fill={rgb,255: red,210; green,105; blue,30}, inner sep=0pt, minimum size=8pt]
\tikzstyle{CirclePurpleSmall}=[draw, shape=circle, fill={rgb,255: red,191; green,0; blue,191}, inner sep=0pt, minimum size=4pt]
\tikzstyle{BigCircleGreen}=[draw, shape=circle, fill={rgb,255: red,0; green,191; blue,0}, inner sep=0pt, minimum size=12pt]
\tikzstyle{BigCircleBlue}=[draw, shape=circle, fill={rgb,255: red,0; green,0; blue,191}, inner sep=0pt, minimum size=12pt]
\tikzstyle{BigCircleRed}=[draw, shape=circle, fill={rgb,255: red,191; green,0; blue,0}, inner sep=0pt, minimum size=12pt]
\tikzstyle{BrownCircleSmall}=[draw, shape=circle,  fill={rgb,255: red,210; green,105; blue,30}, inner sep=0pt, minimum size=6pt]
\tikzstyle{SmallCircleBrown}=[draw, shape=circle,  fill={rgb,255: red,210; green,105; blue,30}, inner sep=0pt, minimum size=5pt]
\tikzstyle{SmallCircleRed}=[draw, shape=circle, fill={rgb,255: red,191; green,0; blue,0}, inner sep=0pt, minimum size=6pt]
\tikzstyle{DashedLine}=[-, densely dashed, line width=0.25mm]
\tikzstyle{DottedLine}=[-, dotted, line width=0.25mm]
\tikzstyle{ThickLine}=[-, line width=0.25mm]
\tikzstyle{ArrowLineRight}=[-, -{Stealth[scale=1.25]}, line width=0.25mm, scale=5]
\tikzstyle{ArrowLineRed}=[-, draw={rgb,255: red,191; green,0; blue,0}, -{Stealth[scale=1.75]}, line width=0.1mm, scale=5]
\tikzstyle{RedLine}=[-, draw={rgb,255: red,191; green,0; blue,0}, fill=none, line width=0.5mm]
\tikzstyle{DashedLineThin}=[-, densely dashed, line width=0.125mm, fill=none, draw=black]
\tikzstyle{DottedRed}=[-, dotted, draw={rgb,255: red,191; green,0; blue,0}, fill=none, line width=0.25mm]
\tikzstyle{DashedRed}=[-, densely dashed, draw={rgb,255: red,191; green,0; blue,0}, fill=none, line width=0.25mm]
\tikzstyle{BlueLine}=[-, draw={rgb,255: red,0; green,0; blue,191}, fill=none, line width=0.5mm]
\tikzstyle{ArrowLineBlue}=[-, draw={rgb,255: red,0; green,0; blue,191}, -{Stealth[scale=1.75]}, line width=0.1mm, scale=5]
\tikzstyle{GreenDoubleArrow}=[<->, draw={rgb,155: red,0; green,255; blue,0},  line width= 0.5mm, scale=5]
\tikzstyle{RedDoubleArrow}=[<->, draw={rgb,255: red,255; green,0; blue,0},  line width= 0.5mm, scale=5]
\tikzstyle{BlueDottedLight}=[-, dotted, draw={rgb,255: red,0; green,0; blue,191}, fill=none, line width=0.3mm]
\tikzstyle{BrownLine}=[-, draw={rgb,255: red,210; green,105; blue,30}, fill=none, line width=0.5mm]
\tikzstyle{DottedRed}=[-, dotted, draw={rgb,255: red,191; green,0; blue,0}, fill=none, dotted, line width=0.5mm]
\tikzstyle{DottedPurple}=[-, dotted, draw={rgb,255: red,191; green,0; blue,191}, fill=none, dotted, line width=0.5mm]
\tikzstyle{BlueDottedLight}=[-, dotted, draw={rgb,255: red,0; green,0; blue,191}, fill=none, line width=0.5mm]
\tikzstyle{ArrowLinePurple}=[-, draw={rgb,255: red,191; green,0; blue,191}, -{Stealth[scale=1.75]}, line width=0.5mm, scale=5]
\tikzstyle{DashedLineGreen}=[-, densely dashed, draw={rgb,255: red,74; green,103; blue,65}, line width=0.25mm]
\tikzstyle{LineGreen}=[-, draw={rgb,255: red, 74; green,200; blue,65}, line width=0.5mm]
\tikzstyle{ArrowLineGreen}=[-, draw={rgb,255: red,0; green,191; blue,0}, -{Stealth[scale=1.75]}, line width=0.5mm, scale=5]
\tikzstyle{GreenLine}=[-, draw={rgb,255: red,0; green,191; blue,0}, fill=none, line width=0.5mm]
\tikzstyle{PurpleLine}=[-, draw={rgb,255: red,191; green,0; blue,191}, fill=none, line width=0.5mm]
\tikzstyle{PPurpleLine}=[-, draw={rgb,255: red,191; green,0; blue,191}, fill=none, line width=2.5mm]
\tikzstyle{DPurpleLine}=[-, dotted, draw={rgb,255: red,191; green,0; blue,191}, fill=none, line width=0.5mm]
\tikzstyle{SBrownLine}=[-, draw={rgb,255: red,191; green,0; blue,191}, fill=none, opacity=0.35, line width=2.5mm]
\tikzstyle{DottedBlue}=[-, dotted, draw=blue, fill=none, dotted, line width=0.5mm]
\tikzstyle{DashedPurpleLine}=[-, densely dashed, draw={rgb,255: red,191; green,0; blue,191}, fill=none, line width=0.5mm]
\tikzstyle{SmallCircleBlue}=[draw, shape=circle, fill=blue, inner sep=0pt, minimum size=5pt]
\tikzstyle{SmallCirclePurple}=[draw, shape=circle, fill={rgb,255: red,191; green,0; blue,191}, inner sep=0pt, minimum size=5pt]
\tikzset{snake it/.style={decorate, decoration=snake}}

\tikzset{
dashstar/.style={
 dash pattern=on 5pt off 5pt,
 postaction={
  decorate,
  decoration={
   markings,
   mark=between positions 9pt and 1 step 10pt with {
     \node[color=red] {*};
   }
  }
 }
},
dashstarstar/.style={ 
 dash pattern=on 5pt off 10pt,
 postaction={
   decorate,
   decoration={
     markings,
     mark=between positions 10pt and 1
          step 15pt
           with {
            \node at (-2pt,0pt) {\pgfuseplotmark{star}};
            \node at (2pt,0pt) {\pgfuseplotmark{star}};
           }
   }
 }
}
}

\begin{document}

\date{February 2026}

\title{Gravitational Background of \\[4mm] Alice-Vortices and R7-Branes}

\institution{PENN}{\centerline{$^{1}$Department of Physics and Astronomy, University of Pennsylvania, Philadelphia, PA 19104, USA}}
\institution{PENNmath}{\centerline{$^{2}$Department of Mathematics, University of Pennsylvania, Philadelphia, PA 19104, USA}}
\institution{Maribor}{\centerline{$^{3}$Center for Applied Mathematics and Theoretical Physics, University of Maribor, Maribor, Slovenia}}

\authors{
Atakan \c{C}avu\c{s}o\u{g}lu\worksat{\PENN}\footnote{e-mail: \texttt{atakanc@sas.upenn.edu}},
Mirjam Cveti\v{c}\worksat{\PENN, \PENNmath, \Maribor}\footnote{e-mail: \texttt{cvetic@physics.upenn.edu}},
\\[0.7em]
Jonathan J. Heckman\worksat{\PENN, \PENNmath}\footnote{e-mail: \texttt{jheckman@sas.upenn.edu}},
Jeffrey Kuntz\worksat{\PENN}\footnote{e-mail: \texttt{kuntzj@sas.upenn.edu}}, and
Chitraang Murdia\worksat{\PENN}\footnote{e-mail: \texttt{murdia@sas.upenn.edu}}
}

\abstract{Codimension-two vortex solutions are important solitonic objects in both quantum field theory and gravity.
In this paper, we construct a class of codimension-two Alice-vortex solutions in axio-dilaton gravity, in which monodromy around the
vortex enacts the axion transformation $C_0 \mapsto -C_0$. In IIB supergravity, this furnishes a class of R7-brane backgrounds
of the sort predicted by the Swampland Cobordism Conjecture. Such configurations generically carry an intrinsic dipole moment. We extract additional properties of such branes from scattering probes. These results provide further evidence that the worldvolume theory of
an R7-brane is an 8D non-supersymmetric interacting quantum field theory.}

\maketitle

\enlargethispage{\baselineskip}

\setcounter{tocdepth}{2}

\tableofcontents

\newpage

\section{Introduction}

Solitons provide an important window into non-perturbative phenomena in quantum field theory (QFT) and gravity.
A particularly striking example is the celebrated $p$-brane solutions of \cite{Horowitz:1991cd} and their relation to the D-branes
of superstring theory \cite{Polchinski:1995mt}. Such objects undergird many key features of quantum gravity.

Especially in the context of string dualities, a core assumption is the existence of sufficient supersymmetry in a given background.
In particular, this means that the candidate brane solutions are automatically BPS and stable. Moreover, at a technical level, it is also far simpler to find explicit solutions since such supersymmetric solutions typically descend to linearized (i.e., first order) differential equations. Finding explicit solutions has led to an immense number of insights in high energy theory, ranging from the discovery of entirely new sorts of quantum field theories (see e.g., \cite{Witten:1995zh, Strominger:1995ac}), to the first explicit examples of the AdS/CFT correspondence \cite{Maldacena:1997re}.

It is natural to ask whether such supersymmetric objects are representative of the class of objects present in quantum gravity.
Indeed, there are also many examples of non-supersymmetric branes, but many of these suffer from string-scale instabilities.\footnote{For a review, see e.g., \cite{Sen:2004nf}.}

Faced with these circumstances, it is natural to resort to more robust techniques which do not require a detailed understanding of the underlying dynamics of the branes. In particular, methods from topology are especially helpful since they are robust against local deformations of a spacetime. Recently, some of these considerations were significantly sharpened in the context of the Swampland Cobordism Conjecture, which asserts that the cobordism group of quantum gravity is trivial \cite{McNamara:2019rup}.\footnote{For recent work on the Swampland Cobordism Conjecture, see e.g., references, \cite{McNamara:2019rup, Montero:2020icj, Dierigl:2020lai, McNamara:2021cuo,
Blumenhagen:2021nmi, Buratti:2021yia, Debray:2021vob, Andriot:2022mri,
Dierigl:2022reg, Blumenhagen:2022bvh, Velazquez:2022eco, Angius:2022aeq,
Blumenhagen:2022mqw, Angius:2022mgh, Blumenhagen:2023abk, Debray:2023yrs,
Dierigl:2023jdp, Kaidi:2023tqo, Huertas:2023syg, Angius:2023uqk,
Kaidi:2024cbx, Angius:2024pqk, Fukuda:2024pvu, Braeger:2025kra,
Heckman:2025wqd, Chakrabhavi:2025bfi, Nevoa:2025xiq, Sen:2025iuf}.}
One striking consequence of this conjecture is that it generically predicts that the spectrum of supergravity must be supplemented by additional \textit{stable} objects \cite{Kaidi:2024cbx, Heckman:2025wqd}. The cobordism conjecture successfully postdicts the existence of objects such as D-branes and orientifold planes, and also leads to striking predictions for non-supersymmetric objects such as domain walls between string vacua (see e.g., \cite{McNamara:2019rup, Heckman:2025wqd}) and reflection 7-branes \cite{Dierigl:2022reg, Dierigl:2023jdp, Chakrabhavi:2025bfi} (see also \cite{Distler:2009ri}) which involve a reflection on the internal torus of F-theory.

But topological considerations alone do not determine basic data such as the metric or stress-energy of the gravitational background. On general grounds, one expects that these branes have an intrinsic singular structure near their core simply because otherwise, they would not be needed to trivialize cobordism defects \cite{Kaidi:2024cbx}. That being said, in the far-field limit, one should expect a smooth background characterized by a low energy gravitational theory. For some recent discussions of the gravitational backgrounds associated to recently discovered non-supersymmetric heterotic branes, see references \cite{Kaidi:2023tqo,Kaidi:2024cbx}.

A tractable setting to make further progress is gravitational backgrounds for Alice-vortices in axio-dilaton gravity. These are codimension-two objects in which the axion $C_0$ transforms as $C_0 \mapsto -C_0$ under monodromy around the brane, i.e., it implements a $\mathbb{Z}_2$ charge conjugation symmetry when acting on this degree of freedom. In the context of the Swampland Cobordism Conjecture, the appearance of this symmetry means we ought to expect a codimension-two defect, namely an Alice-vortex. This can also be viewed as a subsector of type IIB string theory. In that setting, examples of such branes include the $F_L$ and $\Omega$ R7-branes (i.e., reflection 7-branes).\footnote{Let us give some examples of the R7-branes of \cite{Dierigl:2022reg, Dierigl:2023jdp, Chakrabhavi:2025bfi}. Monodromy around the $F_L$ R7-brane of type IIB string theory involves conjugating by left-moving fermion parity, (i.e., $(-1)^{F_L}$) on all fields. We refer to this as the $F_L$ R7-brane. This is a codimension-two vortex-like configuration with the defining property that acts by a sign flip on all of the RR potentials of the theory. There is an S-dual brane known as the $\Omega$ R7-brane which acts by worldsheet orientation reversal. This flips the sign of the RR-potentials $C_0$, $C_4$ and $C_8$ as well as the NSNS potentials $B_2$ and $B_6$. The full transformation also acts non-trivially on fermionic degrees of freedom since the reflection is more properly defined in the $\mathrm{Pin}^{+}$ cover of $\mathrm{GL}(2,\mathbb{Z})$ (see in particular \cite{Chakrabhavi:2025bfi, Pantev:2016nze, Tachikawa:2018njr, Debray:2021vob, Debray:2023yrs}), but these details will play no role in what is to follow. Finally, we comment that reflection branes also arise in IIA and M-theory, as well as more general systems involving an internal reflection symmetry and its accompanying vortex.} For discussion of related Alice-vortex configurations involving a 1-form potential which transforms under charge conjugation, see e.g., \cite{Schwarz:1982ec, Bucher:1991qhl, Bucher:1991bc}. As far as we are aware, however, the explicit gravitational background for these configurations has not appeared in the literature.

Determining the explicit gravitational backgrounds associated with R7-branes would have several ramifications. For one, it would provide strong evidence for their existence. Additionally, having an explicit gravitational background would provide a starting point for a much more detailed analysis. Finally, in the context of holographic backgrounds of the form $\mathrm{AdS}_{D+1} \times X$, an internal reflection on $X$ can be used to build a vortex-like configuration in the $\mathrm{AdS}$ factor, leading to the gravity dual of a charge conjugation symmetry of the boundary CFT$_D$.\footnote{See \cite{Cvetic:2025kdn, Heckman:2025isn, DeMarco:2025pza, Calvo:2025usj, Bah:2025vfu} for related work on the gravity duals of the topological symmetry operators for R-symmetries.}

With these motivations in mind, our aim in this paper will be to determine a class of Alice-vortices of axio-dilaton gravity. In these backgrounds, the main condition we impose is that the axio-dilaton $\tau = C_0 + ie^{-\phi}$ undergoes a monodromy transformation:
\begin{equation}
\tau \mapsto -\overline{\tau}
\end{equation}
in spacetimes which are asymptotically locally flat, and with a non-trivial asymptotic $C_0$ profile. The presence of the vortex generates a conical deficit angle solution, which in turn determines the stress-energy, i.e., the ``tension'' of the brane. We focus on the far-field region, determining a parametric family of solutions for the axio-dilaton and metric. The appearance of a whole family of solutions is due to a few features. First of all, the overall tension is essentially a tunable parameter in the supergravity approximation. Additionally, we find a whole multipole expansion of solutions. This is to be expected because the monodromy $C_0 \mapsto - C_0$ can be accomplished for any odd number of R7-branes.\footnote{We comment that a pair of R7-branes of the same type annihilates to pure radiation \cite{Dierigl:2022reg, Chakrabhavi:2025bfi}.} That being said, we identify particularly ``simple'' solutions which we interpret as a single R7-brane.

A generic feature of our solutions is that there is an intrinsic dipole moment.\footnote{When the $C_0$ profile is forced to be trivial one can entertain solutions with no dipole or higher multi-pole moments.} This is another good indication that the object of interest will not arise from ``standard'' sources in supergravity. Indeed, in classical gravity one typically is restricted to positive mass contributions and as such we can generate monopole and quadrupole contributions; dipoles would require a ``negative tension'' contribution. This is not expected to be much of an issue in string theory backgrounds, since, for example, orientifolds carry both negative RR charge (compared with a BPS D-brane) as well as a negative tension. The presence of the dipole thus points to the appearance of a negative tension source in R7-branes. A related comment is that this is compatible with the fact that the bound state of the R7-branes associated with left-moving fermion parity (the $F_L$ R7-brane) and worldsheet orientation reversal (the $\Omega$ R7-brane) forms a supersymmetric $SO(8)$ 7-brane \cite{Dierigl:2022reg}, i.e., the bound state of $4$ D7-branes and an O7$^{-}$-plane.

In all cases, we find that our class of solutions breaks down as we move towards the core. On general grounds, this is to be expected since we also expect the supergravity approximation to be invalid in this region anyway \cite{Dierigl:2022reg,Heckman:2025wqd}.\footnote{In the case of the heterotic brane solutions, it is possible to give a worldsheet treatment of the near-horizon geometry of the brane \cite{Kaidi:2024cbx}. This luxury is not available for the R7-brane since one expects the branes to be intrinsically strongly coupled near the core \cite{Dierigl:2022reg,Heckman:2025wqd}.} Even so, the appearance of self-consistent asymptotic solutions provides strong evidence for the existence of such objects in gravitational backgrounds. We interpret the breakdown scale as specifying an intrinsic ``thickness'' of the brane which is presumably fully specified in a microscopic formulation.

To gain further insight into the degrees of freedom localized on our brane,
we also consider scalars and pseudo-scalars scattering off of this gravitational background,
i.e., we determine Green's function for these fields. This provides some additional supporting evidence for the conjecture presented in
\cite{Dierigl:2022reg,Heckman:2025wqd} that these branes support interacting degrees of freedom, though we leave a full analysis for future work.

The rest of this paper is organized as follows. In section \ref{sec:SETUP}, we introduce the field equations of axio-dilaton gravity, as well as the boundary conditions and constraints we impose for R7-brane solutions. In section \ref{sec:SOLUTIONS}, we find explicit solutions in the far field regime. Section \ref{sec:SCATTERING} discusses the scattering of scalars and pseudo-scalars off of these gravitational backgrounds.
In section \ref{sec:WORLDVOLUME}, we apply these results to consolidate our understanding of R7-branes. We present our conclusions and potential directions for future work in section \ref{sec:CONC}. Some additional technical details are deferred to the Appendices.

\section{Alice-Vortices and R7-Branes}\label{sec:SETUP}

\begin{figure}[t!]
  \centering
  \begin{tikzpicture}[scale=1.0, line cap=round, line join=round, >=Latex]

  \def\R{1.7}

  \draw[dashed, thick] (0,-0.12) -- (0,{\R+0.20});

  \node[
    star,
    star points=5,
    star point ratio=2.25,
    draw=none,
    fill=red,
    minimum size=7.5mm,
    inner sep=0pt
  ] (vortex) at (0,0) {};

  \def\gap{16}
  \draw[thick,->]
    ({-90+\gap/2}:\R) arc[start angle={-90+\gap/2}, end angle={270-\gap/2}, radius=\R];

  \node[below=3mm] at (0,-\R) {$C_p \mapsto -\,C_p$};

  \begin{scope}[shift={(2.15,1.35)}]
    \draw[thick,->] (0,0) -- (0.8,0) node[below right] {$x$};
    \draw[thick,->] (0,0) -- (0,0.8) node[above left] {$y$};
  \end{scope}

\end{tikzpicture}
  \caption{Monodromy around the $F_L$ R7-brane of type IIB string theory. Here, $x \equiv x^8$ and $y \equiv x^9$ are local coordinates transverse to the R7-brane. The action on the RR potentials is $C_p \mapsto -C_p$. The dashed line indicates a branch cut.}
  \label{fig:monodromy}
\end{figure}
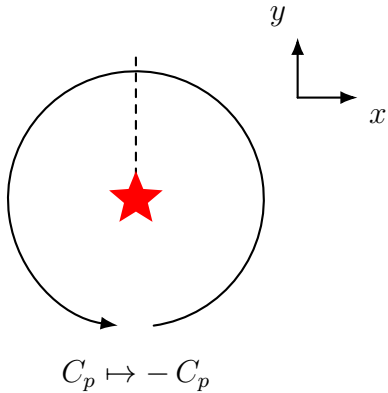

In this section, we lay out some general features of Alice-vortices and the $F_L$ (and $\Omega$) R7-brane of type IIB supergravity. An Alice-vortex is specified by starting with a $D$-dimensional theory with a charge conjugation symmetry $\mathcal{C}$. The main idea is to consider a gauging of this symmetry and vortex configurations such that charge conjugation acts on all fields which wind around this vortex.\footnote{See \cite{Schwarz:1982ec} for a discussion of Alice electrodynamics.} The R7-branes of \cite{Dierigl:2022reg, Debray:2023yrs} amount to a generalization of this picture. In particular, the $F_L$ R7-brane of type IIB string theory is specified by the condition that all RR potentials $C_{p} \mapsto -C_{p}$ after winding around the configuration. This can be viewed as specifying a generalized charge conjugation operation. Since we will keep all higher-form potentials switched off, our results apply to any R7-brane where $C_0 \mapsto -C_0$.\footnote{For the $\Omega$ R7-Brane, the RR potentials $C_2$ and $C_6$ do not flip sign but the NSNS potentials $B_2$ and $B_{6}$ do flip sign. More broadly, there is an entire duality group's orbit of related codimension-two vortices.} See figure \ref{fig:monodromy} for a depiction of an $F_L$ R7-brane.

Geometrically, we can interpret the $(-1)^{F_L}$ action as a reflection on the auxiliary torus of F-theory. Indeed, the axio-dilaton of type IIB supergravity:
\begin{equation}
\tau = C_0 + i e^{-\phi} \,,
\end{equation}
specifies the complex structure modulus of the auxiliary F-theory torus \cite{Vafa:1996xn}. The celebrated IIB dualities correspond to $SL(2,\mathbb{Z})$ transformations of the form $\tau \mapsto (a \tau + b) / (c \tau + d)$ with:
\begin{equation}
\begin{bmatrix}
        a & b \\ c & d
    \end{bmatrix} \in \mathrm{SL}(2,\mathbb{Z})\,.
\end{equation}
Including the contributions from left-mover fermion parity (namely $(-1)^{F_L}$) and orientation reversal (namely $\Omega$) extends this to $\mathrm{GL}(2,\mathbb{Z})$, and the further extension to include the action on fermions takes us to the $\mathrm{Pin}^{+}$ cover of $\mathrm{GL}(2,\mathbb{Z})$.\footnote{See in particular \cite{Chakrabhavi:2025bfi, Pantev:2016nze, Tachikawa:2018njr, Debray:2021vob, Debray:2023yrs}.} As an element of $\mathrm{GL}(2,\mathbb{Z})$, the symmetry $(-1)^{F_L}$ acts as $\mathrm{diag}(-1,1)$, i.e., it is an orientation reversing reflection of the F-theory torus. Under monodromy, the axio-dilaton transforms as:
\begin{equation}
\tau \mapsto - \overline{\tau}\,.
\end{equation}
Topologically, the F-theory torus is just an $S^1_{A} \times S^1_{B}$. Fibering this over an angular direction $S^{1}_{\infty}$ of the 10D spacetime far from the brane, this winding around the vortex constructs $\mathrm{KB_{\infty}} \times S^{1}_{B}$, i.e., a Klein Bottle with a spectator $S^{1}_{B}$ (the circle which is stationary under the reflection. We can extend this geometry in the radial direction transverse to the R7-brane and thus produce a conical geometry $\mathrm{Cone}(\mathrm{KB} \times S^{1})$, in the obvious notation.

Our aim will be to construct the gravitational backgrounds compatible with these topological constraints. To focus on the essential features, we will restrict our attention to IIB backgrounds with a non-trivial axio-dilaton and metric profile, i.e., we keep all other $p$-form potentials switched off. One can in principle study a more general class of backgrounds with appropriate monodromy (i.e., $C_{p} \mapsto - C_{p}$) but we defer this to future work. Additionally, we assume that our metric is asymptotically locally flat as we move far away from the brane. Taking into account these simplifying assumptions, it will therefore suffice to consider solutions to axio-dilaton gravity. After spelling out the general equations of motion for this system, we discuss some general properties of such solutions. In section \ref{sec:SOLUTIONS}, we solve the corresponding equations of motion for an Alice-vortex / R7-brane.

\subsection{Axio-Dilaton Gravity}

The axio-dilaton system coupled to gravity is characterized by the action:
\begin{equation} \label{axiodilaton_action}
    S = \frac{1}{2\kappa^2}\int \dd[10]{x} \sqrt{-g} \left(R - \frac{\partial_\mu \tau \partial^\mu \bar{\tau}}{2 (\Im\tau)^2} \right) \,.
\end{equation}
where $\tau = C_0 + i e^{-\phi}$ is the axio-dilaton field. We shall be interested in solutions to the Einstein field equations:
\begin{equation}
G_{\mu \nu } = \kappa^2 T_{\mu \nu}\,,
\end{equation}
with $G_{\mu \nu} $ the Einstein tensor and $T_{\mu \nu }$ the stress-energy tensor sourced by the axio-dilaton profile. In what follows, we work in units where $\kappa^{2} = 1$, i.e., we solve $G_{\mu \nu} = T_{\mu \nu}$. See Appendix \ref{app:DDIM} for a discussion of axio-dilaton gravity in $D$-dimensions.

The equation of motion obtained by varying the action in \eqref{axiodilaton_action} with respect to $\bar{\tau}$ is
\begin{align} \label{EOMtauFull}
    \frac{1}{\sqrt{-g}} \partial_\mu \left(\sqrt{-g} g^{\mu \nu} \partial_\nu \tau \right) + i\frac{\partial_\mu \tau \partial^\mu \tau}{\Im\tau} = 0 \,,
\end{align}
while variation with respect to $\tau$ yields the complex conjugate equation governing $\bar{\tau}$.
Lastly, the Einstein field equations for our system can be obtained by varying with respect to the metric:
\begin{align} \label{EOMgravitational}
    G_{\mu \nu} - \frac{1}{4(\Im\tau)^2}\left(\partial_\mu \tau \partial_\nu \bar \tau + \partial_\nu \tau \partial_\mu \bar \tau - g_{\mu \nu}\partial_\rho\tau\partial^\rho\bar{\tau} \right) = 0 \,.
\end{align}

Let us now specialize to codimension-two objects, i.e., 7-branes. A celebrated class of solutions which preserve supersymmetry are the cosmic string backgrounds of \cite{Vafa:1996xn,Greene:1989ya}. An important feature of such solutions is that in the directions transverse to the brane, the profile of the axio-dilaton is manifestly holomorphic, i.e., $\tau(z)$ for $z$ a local holomorphic coordinate transverse to the brane. A very helpful feature of retaining supersymmetry is that solving the field equations only requires dealing with first order differential equations.

In the case of R7-branes, however, such luxuries are unavailable. Indeed, these backgrounds break all supersymmetries \cite{Dierigl:2022reg, Chakrabhavi:2025bfi} and so we must proceed more generally. Indeed, the anti-periodic monodromy condition
for the $C_0$ potential translates to the following condition for the axio-dilaton:
\begin{equation} \label{tauMonodromy}
    \tau(r,\theta+2\pi) = -\bar{\tau}(r,\theta) \,,
\end{equation}
where $r$ is the radial direction of the brane and $\theta$ is an angular coordinate which winds around it. We shall focus on the generic situation where asymptotically, $C_0$ is non-zero. Manifestly, the metric is periodic in $\theta$. We will be interested in non-trivial codimension-two vortex solutions to the equations of motion that satisfy these constraints. 

Since we are interested in a solution corresponding to a codimension-two defect, we separate the 10D spacetime metric into the 8-dimensional worldvolume and 2-dimensional transverse components by writing
\begin{align}
	\dd{s}^2 = g_{\mu \nu} \dd{x}^\mu \dd{x}^\nu &= g_{ab} \dd{x}^a \dd{x}^b +  g_{ij} \dd{x}^i \dd{x}^j \,,
\end{align}
where Greek indices $\mu$, $\nu$, ...\ refer to the full 10D metric components, early-alphabet Latin indices $a$, $b$, ...\ refer to the 8D external worldvolume of the brane (including the time dimension $x^0 = t$), and mid-alphabet Latin indices $i$, $j$, ...\ refer to the 2D internal (transverse) space. This ansatz is motivated by the arguments in \cite{Kaidi:2024cbx, Heckman:2025wqd} that R7-branes are actually stable objects in supergravity. Additionally, the form of our solution assumes 8D Lorentz invariance, i.e., there is no angular momentum in the configuration.\footnote{It would of course be interesting to consider more general solutions with angular momentum included.}

Due to translational and rotational invariance along the worldvolume of the brane, we expect the metric of the 8D worldvolume $g_{ab}$ to have the same symmetries as Minkowski space, and it will therefore be locally flat up to a warp factor that depends only on the 2D transverse coordinates. Additionally, the metric of the 2D transverse space $g_{ij}$ can always be put into a conformally flat form by some coordinate transformation. Therefore, we consider the ansatz:
\begin{equation} \label{metricAnsatz}
    \dd{s}^2 = e^{f(x^i)}\eta_{ab}\dd{x}^a\dd{x}^b + e^{h(x^i)} \delta_{ij}\dd{x}^i\dd{x}^j \, ,
\end{equation}
where $\eta_{ab}$ is the 8D Minkowski metric, $\delta_{ij}$ is the 2D Euclidean metric, and $f(x^i)$ and $h(x^i)$ are arbitrary functions parameterizing two independent warp factors that depend only on the coordinates $(x^8,x^9)$ in the 2D transverse space. We also assume that $\tau$ depends only on the two transverse coordinates.

With this ansatz, one finds that a sum of the $\mu= \nu = 8$ and $\mu=\nu=9$ Einstein's equations in \eqref{EOMgravitational} yields the simple condition:
\begin{align} \label{fcon}
	\partial_i\partial_if + 4 \partial_if \partial_if = 0 \,,
\end{align}
where repeated lower indices indicate a sum with respect to the flat space Euclidean metric on the 2D transverse space. This equation can be further simplified by redefining the warp factor as $f = \frac{1}{4}\log F$, leading to a 2D Laplace's equation:
\begin{align}
    \partial_i\partial_iF = 0 \,.
\end{align}
In other words, $F(x^i) = e^{4 f(x^i)}$ must be a 2D harmonic function.

The remaining equations can be more easily expressed in holomorphic and antiholomorphic coordinates. For this purpose, we define the coordinates $z \equiv x^8 + i x^9$ and $\bar{z} \equiv x^8 - i x^9$. These definitions correspond to the metric
\begin{align}
	\dd{s}^2 = e^{f(z,\bar{z})}\eta_{ab}\dd{x}^a\dd{x}^b + e^{h(z,\bar{z})}\dd{z}\dd{\bar{z}} \,.
\end{align}
In these coordinates, the equation of motion in \eqref{EOMtauFull} for the axio-dilaton $\tau$ takes the form
\begin{align} \label{tauEOM}
	\partial\bar{\partial}\tau + 2\big(\partial f\bar{\partial}\tau + \bar{\partial} f\partial\tau\big) + i\frac{\partial\tau\bar{\partial}\tau}{\Im\tau} = 0 \,,
\end{align}
where $\partial \equiv \partial_z$ and $\bar{\partial} \equiv \partial_{\bar{z}}$. The Laplace's equation for $F$ simply becomes $\partial \bar{\partial} F = 0$ or equivalently $\partial \bar{\partial} f + 4 \partial f \bar{\partial} f = 0$.

As we show in Appendix \ref{app:DDIM}, in addition to Laplace's equation $\partial \bar{\partial}F=0$, the complete set of Einstein's equations simply reduces to three other independent equations. Concretely, the $G_{ab} = T_{ab}$, $G_{zz} = T_{zz}$, and $G_{\bar{z} \bar{z}} = T_{\bar{z} \bar{z}}$ equations respectively yield
\begin{align} \label{ijEOM}
	\partial\bar{\partial} h + \frac{7}{2}\,\partial\bar{\partial} f = -&\frac{\partial\tau\bar{\partial}\bar{\tau} + \bar{\partial}\tau\partial\bar{\tau}}{4(\Im\tau)^2} \, , \\
    \label{abEOM1}
    \partial^2 f + \frac12(\partial f)^2 - \partial f \partial h &= -\frac{\partial \tau \partial \bar{\tau}}{8(\Im\tau)^2} \, , \\
    \label{abEOM2}
    \bar{\partial}^2 f + \frac12(\bar{\partial} f)^2 - \bar{\partial}f\bar{\partial}h &= - \frac{\bar{\partial} \tau \bar{\partial} \bar{\tau}}{8(\Im\tau)^2} \,.
\end{align}

It is worth mentioning that these three equations are not completely independent after including the equation of motion for $\tau$. To see this explicitly, we implicitly introduce $h_0$ via:
\begin{align} \label{hredef}
	h(z,\bar{z}) = h_0(z,\bar{z}) + \frac92f(z,\bar{z}) + \log\big(\partial f\bar{\partial} f\big) \, ,
\end{align}
which allows \eqref{abEOM1} and \eqref{abEOM2} to be expressed as
\begin{equation} \label{abEOMswithh0}
    \partial f \, \partial h_0 = \frac{\partial \tau \, \partial \bar{\tau}}{8(\Im\tau)^2} \, , \qquad \bar{\partial} f \, \bar{\partial} h_0 = \frac{\bar{\partial} \tau \, \bar{\partial} \bar{\tau}}{8(\Im\tau)^2}\, .
\end{equation}
Taking (anti-)holomorphic derivatives of these equations and using the equations of motion for $\tau$ and $\bar{\tau}$ allows us to derive
\begin{align} \label{h0eq}
	\partial f\left(\partial\bar{\partial} h_0 + \frac{\partial\tau\bar{\partial}\bar{\tau} + \bar{\partial}\tau\partial\bar{\tau}}{4(\Im\tau)^2}\right) = 0 \,, \qquad \bar{\partial} f\left(\partial\bar{\partial} h_0 + \frac{\partial\tau\bar{\partial}\bar{\tau} + \bar{\partial}\tau\partial\bar{\tau}}{4(\Im\tau)^2}\right) = 0 \,.
\end{align}
Rewriting this equation in terms of $h$ using \eqref{hredef}, we recover \eqref{ijEOM}, demonstrating that we do not have independent equations of motion as long as $f$ is not a constant.

Before we proceed to solving this set of equations, there are two important things to note:
\begin{itemize}
    \item The warp factor $f$ is constant if and only if $\tau$ is holomorphic or antiholomorphic, i.e., $\tau$ depends only on either $z$ or $\bar{z}$. This can be seen from equations \eqref{tauEOM} and \eqref{abEOM1}. Indeed, \eqref{tauEOM} together with the reality condition on $f$ implies that $\partial f = \bar{\partial} f = 0$ if either $\partial \tau = 0$ or $\bar{\partial} \tau = 0$.\footnote{Restricting $f$ to be real allows us to conclude that $\partial f =0 \iff \bar{\partial}f = 0$.} Moreover, \eqref{abEOM1} implies that $\partial \tau = 0$ or $\bar{\partial} \tau = 0$ if $f = \text{constant}$. It is not possible to find a non-trivial holomorphic solution for $\tau$ that satisfies the monodromy condition in \eqref{tauMonodromy}, which forces us to consider a non-trivial $f$ for our solutions. However, holomorphic solutions with different monodromy conditions can still be of interest, an example of which is the so-called ``stringy cosmic string'' solution, investigated in \cite{Greene:1989ya} (see also \cite{Vafa:1996xn}).
    \item Under a holomorphic transformation $z \rightarrow z^{\prime}(z)$, $\bar{z} \rightarrow \bar{z}^{\prime}(\bar{z})$, all equations of motion remain invariant up to an overall scale, with the functions $f$, $h_0$, and $\tau$ not transforming,\footnote{To be precise, $f$, $h_0$, and $\tau$ transform like scalars as $f(z,\bar{z}) \to f^\prime(z,\bar{z}) = f(z^\prime(z),\bar{z}^\prime(\bar{z}))$, $h_0(z,\bar{z}) \to h_0^\prime(z,\bar{z}) = h_0(z^\prime(z),\bar{z}^\prime(\bar{z}))$, and $\tau(z,\bar{z}) \to \tau^\prime(z,\bar{z}) =\tau(z^\prime(z),\bar{z}^\prime(\bar{z}))$.} as can be seen from equations \eqref{fcon}, \eqref{tauEOM}, and \eqref{abEOMswithh0}. However, $h(z,\bar{z})$ transforms like
    \begin{equation}
        h(z,\bar{z}) \to h^\prime(z,\bar{z}) = h(z^\prime(z),\bar{z}^\prime(\bar{z})) + \log\left(\dv{z^\prime(z)}{z} \dv{\bar{z}^\prime(\bar{z})}{\bar{z}}\right)
    \end{equation}
    due to the transformation of the $\partial^2f$ and $\bar{\partial}^2 f$ terms in \eqref{abEOM1} and \eqref{abEOM2}. Therefore, if one finds any solution $(f,h,\tau)$ to the equations of motion presented in this section, it is possible to obtain an infinite-parameter class of solutions $(f^\prime, h^\prime, \tau^\prime)$ by simply performing holomorphic transformations on the solution $(f,h,\tau)$. In the next section, we will interpret this conformal freedom as providing a simple way to start with a single seed solution and generate other solutions with different sets of multipole moments in the directions transverse to the brane.
\end{itemize}

\section{Asymptotic Solutions} \label{sec:SOLUTIONS}

Having introduced the equations of motion and monodromy conditions for Alice-vortices in axio-dilaton gravity, we now proceed to construct explicit solutions. We proceed via the method of power series solutions in the far field limit. A generic feature of our solutions is the presence of a dipole in the energy-momentum sourced by the brane. This can in principle be removed by tuning the solution, but we always have a higher-order multipole moment.

\subsection{Power Series Solutions}

We begin by finding solutions for the warp factor $f$ through the requirement $\partial_i \partial_i F = 0$. Since its Laplacian vanishes, $F(z,\bar{z})$ must be the sum of a holomorphic and an antiholomorphic function, $F(z,\bar{z}) = u(z) + v(\bar{z})$. Requiring $F$ to be real, we obtain
\begin{equation}
    F(z,\bar{z}) = \mathcal{F}(z) + \overline{\mathcal{F}}(\bar{z}) \,, \qquad\qquad f(z,\bar{z}) = \frac{1}{4} \log\Big(\mathcal{F}(z) + \overline{\mathcal{F}}(\bar{z})\Big) \,,
\end{equation}
for any holomorphic function $\mathcal{F}(z)$.

Moreover, we require that $f$ approaches zero as $r\to\infty$ (or $z \to \infty$) and that $f$ is periodic in $\theta$. The far-field expansion for the most general solution for large $z$ satisfying these constraints is
\begin{equation} \label{Fseries}
    F(z,\bar{z}) = 1 +\sum_{k=1}^{\infty} \left( a_k z^{-k} + \bar{a}_k \bar{z}^{-k} \right) \, .
\end{equation}
Using the conformal freedom associated with the solutions, we can first drop all higher-order terms except for $a_1$ and search for a solution with
\begin{equation}
\label{fSolz}
    f(z,\bar{z}) = \frac{1}{4} \log\big(F(z,\bar{z})\big) = \frac{1}{4}\log(1 + az^{-1} + a \bar{z}^{-1}) \, .
\end{equation}
We can choose $a > 0$ as the phase of $a$ can be eliminated by a rotation of the $z$-plane. It is also possible to set $a = 1$ by an appropriate rescaling of the $z$-plane, however, as $a$ is a dimensionful quantity, we prefer to keep it arbitrary. After finding a solution with this assumption, a general solution can be obtained by a holomorphic transformation $z \to z^{\prime}(z)$.

We can also express these functions in polar coordinates $(r,\theta)$
\begin{align} \label{fSol}
	F(r,\theta) = 1 + \frac{2a \cos\theta}{r} \,,
    \qquad \qquad
	f(r,\theta) = \frac14\log(1 + \frac{2a \cos\theta}{r}) \,.
\end{align}
Evidently, this solution is only valid in the region $r>2a$, outside of which the metric becomes complex. We will return to studying the equations of motion near this region later on.

It is also worth mentioning that periodic solutions to Laplace's equation also include the logarithmic term
\begin{align}
    F \supset a_0 \log \big( z\bar{z} \big) = 2 a_0 \log r \, .
\end{align}
However, this term, which corresponds to a monopole source transverse to the brane, is forbidden by requiring the bulk warp factor in the directions parallel to the brane to be finite as $r\to\infty$.  The term that persists in \eqref{fSol} is a dipole contribution.\footnote{As already noted in the Introduction, the appearance of a dipole moment does not arise in classical gravity, since it would require a negative tension object. This is less of an issue in string theory, where negative tension objects such as orientifold planes are already available. Indeed, the bound state of a $F_L$ and $\Omega$ R7-brane forms an $\mathrm{SO(8)}$ 7-brane, i.e., a BPS configuration with four D7-branes and an O7$^{-}$-plane. A further comment is that one can actually tune our solution so that the multipole expansion starts at some other higher order. We do not expect these highly tuned situations to characterize the R7-brane.}

Since the equations for  $h$ and $\tau$ are more involved, finding closed-form solutions for them is not as tractable as it was for $f$. However, it is possible to find power series solutions to these functions that are valid in the same $r>2a$ region where $f$ is well-defined. Thus, we search for asymptotic solutions to \eqref{tauEOM}--\eqref{abEOM2} starting from the series ans\"{a}tze
\begin{align}
    h(z,\bar{z}) &= -b\log(z\bar{z}) + \sum_{m,n=0}^\infty b_{m,n}\left(\frac{z}{a}\right)^{-m/2}\left(\frac{\bar{z}}{a}\right)^{-n/2} \,, \\
    \tau(z,\bar{z}) &= \lambda \sum_{m,n=0}^\infty c_{m,n}\left(\frac{z}{a}\right)^{-m/2}\left(\frac{\bar{z}}{a}\right)^{-n/2} \,,
\end{align}
where $\lambda$ is an arbitrary real constant that accounts for the equations of motion being invariant under an overall scaling of $\tau$. Furthermore, we have included a logarithmic term in the expansion for $h$, with $b > 0$, in anticipation of the presence of a conical defect.

Before solving the equations of motion, we first enforce the appropriate monodromy conditions.
Requiring that $h$ is periodic under a $2\pi$-rotation, we find
\begin{align}
	\big((-1)^{m+n} - 1\big)b_{m,n} = 0 \,,
\end{align}
which implies that all the $b_{m,n}$ with $m+n=\text{odd}$ must vanish. Furthermore, requiring $h$ to be real forces the coefficients $b_{m,n}$ to be Hermitian:
\begin{equation}
    \overline{b}_{m,n} = b_{n,m} \, .
\end{equation}
Similarly, the monodromy condition, $\tau(r,\theta+2\pi) = -\bar{\tau}(r,\theta)$ or $\tau(e^{2 \pi i}z, e^{-2\pi i}\bar{z}) = -\bar{\tau}(z,\bar{z})$,  gives
\begin{align}
	\overline{c}_{m,n} = (-1)^{m+n+1}c_{n,m} \,.
\end{align}
Finally, we can set $b_{0,0} = 0$ since a constant term in $h$ corresponds to an irrelevant overall scale for the metric and does not appear in the equations of motion. We can also set $c_{0,0} = i$ since an arbitrary rescaling of $\tau$ is already accounted for by $\lambda$.

Inserting all of the above into the equations \eqref{tauEOM}--\eqref{abEOM2}, along with the series expansion of $f$ in \eqref{fSol},
\begin{align}
	f(z,\bar{z}) &= \frac14\log(1 + az^{-1} + a\bar{z}^{-1}) \nonumber\\
	&= \frac{a}{4}(z^{-1}+\bar{z}^{-1}) - \frac{a^2}{8}\left(z^{-2} + 2z^{-1}\bar{z}^{-1} + \bar{z}^{-2}\right) + \cdots \,,
\end{align}
allows one to solve the equations order by order and find the solutions for $h$ and $\tau$
{\allowdisplaybreaks
\begin{align}
	h(z,\bar{z}) =& -b\log(z\bar{z}) - 4(2-b)az^{-1/2}\bar{z}^{-1/2} \nonumber\\
    & +\left[\bigg(3i\sqrt{\frac{2 - b}{2}}c_3 - \frac{c_2^2}{2} - (2 - b)\big(1 + 2ic_2\big) - \frac{7}{8}\bigg)az^{-1} + \text{c.c.}\right]+ \cdots \, , \label{hSol} \\
	\Re\tau(z,\bar{z}) =& \pm \lambda\bigg(2i\sqrt{2(2-b)}a^{1/2}z^{-1/2} + c_3\,a^{3/2}z^{-3/2} \nonumber\\
    & \quad + i\sqrt{2(2-b)}\left( 8(2-b) - 2ic_2 -1 \right) a^{3/2}z^{-1}\bar{z}^{-1/2} + \cdots + \text{c.c.}\bigg) \, , \label{RetauSol} \\
	\Im\tau(z,\bar{z}) =& \lambda \bigg(1 - ic_2az^{-1} + i \overline{c}_2 a \bar{z}^{-1} + 8(2-b)az^{-1/2}\bar{z}^{-1/2} + \cdots \bigg) \, . \label{ImtauSol}
\end{align}}
It should be noted that, in addition to the unfixed constants $a$ and $b$, we find additional unfixed complex constants $c_i$ (the coefficients $c_i \equiv c_{i,0}$ appearing in the series expansion of $\tau$) which appear at every higher order in our solutions. We thus see how the redundancies in Einstein's equations concretely affect the series solutions. At least some of these redundancies can be fixed by examining the behavior of this solution near the origin.

\subsection{Multipole Sources on the Brane} \label{sec:OriginSols}

We have already found solutions to the vacuum equations in the large $r$ regime, but we have not yet analyzed the behavior near the potential singularity at $r=0$. Although our solutions are only valid away from the brane ($r\gtrsim a$), we can gain some insight about the small $r$ region by using Gauss's Law and the divergence theorem. Singular terms near the origin generically generate delta functions (and their derivatives) which can be accounted for by introducing a sum of multipole moments at the brane. Therefore, we can write the total energy-momentum tensor $T_{\mu \nu}$ as the sum of the non-singular bulk contribution sourced by the axio-dilaton $\tau$, $T^{(\text{bulk},\tau)}_{\mu\nu}$, and the singular contribution sourced by the brane, $T^{(\text{brane})}_{\mu\nu}$, by writing $T_{\mu\nu} = T^{(\text{bulk},\tau)}_{\mu\nu} + T^{(\text{brane})}_{\mu\nu}$ with
\begin{align}
	&T^{(\text{brane})}_{ab} = -2\pi g_{ab}\Big(\mu\,\delta^{(2)}_g(\vec{r}) -  Q^i\nabla_i\delta^{(2)}_g(\vec{r})  + \cdots\Big) \, , \\
    &T^{(\text{brane})}_{ij} = 2\pi g_{ij}\Big(q^i\nabla_i\delta^{(2)}_g(\vec{r})  + \cdots\Big) \,, \label{EMsources}
\end{align}
where $\delta^{(2)}_g(\vec{r}) $ is the curved space delta function that satisfies $\int\dd[2]{x}\sqrt{g_\perp}\delta^{(2)}_g(x)=1$ on the transverse space. Here, $\mu$ is the brane tension (monopole moment) and $Q^i$, $q^i$ are dipole moments sourced by unspecified degrees of freedom living on the brane. The ellipses represent higher multipole sources that we do not explicitly consider here since it is straightforward to generalize the following analysis to these higher moments. This ansatz for the energy-momentum tensor is justified by considering the functional forms of $f$ and $h$, and attributing delta function terms to jump discontinuities in the Laplacians of $f$ and $h$ at $r=0$. The relevant expression in the Einstein's equations, the traces over the worldvolume and transverse spaces, are respectively given by
\begin{align}
	&e^{-h}\bigg(\Delta f + \frac17\Delta h\bigg) + (\text{regular}) = -\frac{4\pi}{7}\Big(\mu\,\delta^{(2)}_g(\vec{r})  - Q^i\nabla_i\delta^{(2)}_g(\vec{r}) \Big) + \cdots \, ,\label{aaEq} \\
	&e^{-h}\Delta f + (\text{regular}) = \pi q^i\nabla_i\delta^{(2)}_g(\vec{r})  + \cdots \,. \label{iiEq}
\end{align}

Each of the equations above can be related to a Gauss's law-type flux via the divergence theorem, which can then be equated to a straightforward integral of the delta function sources. The details of this calculation can be found in Appendix \ref{app:GL} and we state just the final results here. First, the brane tension is identified with $b$, the coefficient of the leading logarithmic term in $h$
\begin{align} \label{bDef}
	\mu = b \,,
\end{align}
as expected because this term corresponds to a defect-sourcing monopole that lies along the brane. The transverse dipole moments are given by the arbitrary constant appearing in the solution for $f$,
\begin{align} \label{aDefs}
	q_x = a \,, \qquad\qquad q_y = 0 \,,
\end{align}
where, to avoid overloading the notation, we have introduced $x \equiv x^8$ and $y \equiv x^9$. Without loss of generality, we have also presented the dipole solution for pointing along just the $x$-axis of the 2D plane transverse to the brane.\footnote{This amounts to a spontaneous symmetry breaking of the isometries transverse to the brane. See section \ref{sec:WORLDVOLUME} for additional discussion.}
The longitudinal dipole moments are related to the $c_i$'s and the transverse dipole moments,
\begin{align} \label{Qcons}
	Q_x + q_x u(c_2,c_3) + q_y v(c_2,c_3) = 0 \,, \qquad\qquad Q_y - q_x v(c_2,c_3) + q_y u(c_2,c_3) = 0 \,,
\end{align}
where $u$ and $v$ are real-valued functions of the $c_i$'s.

\begin{figure}[t]
  	\centering
	\begin{tikzpicture}[scale=1]


  \draw[fill=gray!20] (0,2) ellipse (0.2 and 0.07);

  \draw[fill=gray!20] (-0.2,-2) -- (-0.2,2)
                      arc (180:360:0.2 and 0.07)
                      -- (0.2,-2)
                      arc (360:180:0.2 and 0.07);

  \draw[<->] (0,-2.3) -- (0,2.25) node[above right] {$x^a$};

  \draw (0,0) ++(0:1.6 and 0.6) arc (0:167:1.6 and 0.6);
  \draw (0,0) ++(193:1.6 and 0.6) arc (193:360:1.6 and 0.6);

  \draw[->] (0,0) -- (1.6,0) node[above right] {$r$};

  \draw[->] (1.6,0) arc (0:40:1.6 and 0.6);
  \node at (1.3,0.65) {$\theta$};

  \draw[thick] (0,0) -- ({1.6*cos(167)},{0.6*sin(167)});
  \draw[thick] (0,0) -- ({1.6*cos(193)},{0.6*sin(193)});

  \draw (0,0) ++(175:0.72) arc (175:185:0.72);
  \node at (-0.72,0.25) {$\delta$};

  \draw (0,-2.5) -- (0.2,-2.5);
  \draw (0,-2.55) -- (0,-2.45);
  \draw (0.2,-2.55) -- (0.2,-2.45);
  \node at (0.1,-2.75) {$r_c$};

  \node at (0,-3.4) {global geometry};


  \begin{scope}[shift={(8,0)}, scale=1.62]

    \draw (0,0) circle (1.2);

    \begin{scope}
      \clip (0,0) circle (1.2);
      \shade[left color=blue!80, right color=red!80]
        (-1.2,-1.2) rectangle (1.2,1.2);
    \end{scope}

    \draw[->] (0,0) -- (0.9,0) node[pos=1,above] {$r$};
    \draw[->] (0.65,0) arc (0:35:0.65) node[pos=1,above] {$\theta$};

    \fill (0,0) circle (1.2pt);
    \node at (0,-0.3) {$\mu$};

    \node at (0,-2.1) {generic dipole source profile};

  \end{scope}

\end{tikzpicture}
  	\label{fig:rod-multipole}
	\caption{
	\textit{Left:} Geometry of the R7--brane, shown as a codimension--2 source with core radius $r_c<r_0$. The transverse plane has a deficit angle $\delta$.
	\textit{Right:} Illustrative sample internal structure of the brane stress--energy; varying stress-energy that generates the dipole moments is indicated by the gradient.}
\end{figure}
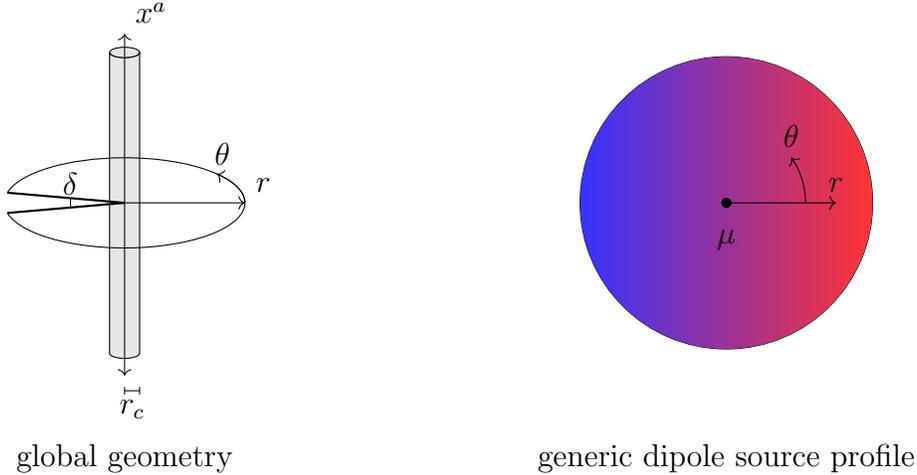

Since the warp factor $f$ is non-trivial only for $a\neq0$, we can conclude that $\tau$ has a non-trivial monodromy only if $a$ is non-zero, which is equivalent to the transverse dipole moment in the transverse components of the energy-momentum tensor being non-zero.\footnote{Setting $a = 0$ along with all higher order terms is in principle possible, but means that $\tau$ is exactly constant and so (to satisfy the monodromy constraint) has trivial $C_0$ profile. One can in principle entertain such solutions, but we emphasize that our focus here is on the generic situation with non-zero $C_0$.} The same conclusion does not hold for $Q_x$ and $Q_y$, the dipole moments in the longitudinal components of the energy-momentum tensor. These moments can take any value, including zero, without affecting the qualitative nature of our solution, i.e., without spoiling the monodromy, trivializing the $\tau$ solution, etc. Higher-order calculations suggest that this behavior holds for the higher multipole moments in $T_{ab}$ as well.

For simplicity, we thus consider a source structure where the dipole moments in $T_{ab}$ vanish, and we restrict to the case with a $\tau$ profile that generates only a monopole source term in the longitudinal components of the energy-momentum tensor. This is quantified by the condition $u(c_2,c_3) = v(c_2,c_3) = 0$, which is solved by
\begin{align}
	c_3 = -\frac{i}{12\sqrt{2(2-\mu)}}\big(4c_2^2 - 7\big) + \frac{\sqrt{2-\mu}}{3\sqrt{2}}(4c_2 - 2i) \,.
\end{align}
This condition also fixes the $h$ expansion in \eqref{hSol} to be
\begin{align}
	h(z,\bar{z}) = -\mu \log(z\bar{z}) - \frac{7}{4}qz^{-1} - \frac{7}{4}\bar{q}\bar{z}^{-1} + 4(2-\mu) \sqrt{q \bar{q}}z^{-1/2}\bar{z}^{-1/2} + \cdots \label{hSolfixed} \,,
\end{align}
where we define $q = q_x + i q_y$. Although \eqref{aDefs} implies that $q = a$, we prefer the above form because we have replaced the arbitrary parameter $a$ by the dipole moment $q$, which is a physical quantity. Thus, by assuming that $T_{ab}$ only has a monopole source centered at the origin, we are able to fix half of the remaining $c_i$'s in the $\tau$ expansion, while also fixing the warp factor solutions entirely in terms of physical data.

We can give a similar interpretation of the higher order terms appearing in \eqref{fSol} as well as those in the power series expansion of $h$.
These contributions specify higher-order multipole moments of the brane, so that a $2^n$-pole contributes terms to the energy-momentum tensor of the schematic form $T^{(\mathrm{brane})}_{ij} \supset q^{n}\partial^{n} \delta_{g}(\overrightarrow{g})$. Indeed, one way to generate examples of this sort is to first begin with the background with a leading order dipole moment in the transverse direction. Under the holomorphic map $z \mapsto z^{n}$, we generate a new solution, and the monopole term shifts as $\mu \mapsto n \mu$. The original dipole is then replaced by a leading order $2^n$-pole configuration. For more general holomorphic maps $z \mapsto \psi(z)$, the zeros of $\psi(z)$ specify the locations of the monopole sources. An analysis of the next few higher orders in our solutions indicates that the higher longitudinal moments (components of $T_{ab}^{(\mathrm{brane})}$)can likely also be set to zero by tuning the new arbitrary constants that appear at each order in the asymptotic series expansion.

With the identification of $a$ as the transverse dipole source, we can also make statements about the radius of convergence of our solutions. Observe that $h$ and $\tau$ are essentially series expansions in $q/r$. Since the $c_i$'s are $\order{1}$, we can expect the expansions to converge in the region $r > r_0$ with $r_0 \sim a$ up to some numerical factor. In most settings with physical relevance, we expect that the components of the dipole moment, particularly in the transverse direction, are significantly smaller than the monopole coefficient, which is itself $\order{1}$. This is because the monopole term corresponds to the brane tension, whereas the dipole term reflects microscopic asymmetry. We can therefore infer that our asymptotic series solutions are valid over a wide range of $r$, at least up to the region where we can begin to resolve the microscopics that source the dipole moment. Evidently, this nearly matches the region where the solution for $f$ is valid.

\subsection{Conical Deficit Angle}

In the previous subsection, we have confirmed the presence of a brane with non-zero tension (i.e., the monopole term) at the origin of the 2D transverse space. Thus, it is reasonable to expect a non-trivial holonomy from winding around the origin. In particular, since the transverse warp factor in \eqref{hSolfixed} goes like $h \approx - 2\mu \log r$ at large $r$, the coordinate transformation $\rho\propto r^{1-\mu}$ brings the transverse metric into the form $\dd{\rho}^2+(1-\mu)^2\rho^2\dd{\theta}^2$ which is the metric for a cone with deficit angle $\delta = 2\pi\mu$.

To derive this result more concretely, we consider the parallel transport of a vector $V$ around a curve $\gamma$ in the transverse space. Expressing this vector as $V=V^ie_i$, where $e_j$ is an orthonormal frame on the transverse space ($i,j\in\{8,9\}$), the parallel transport equation is
\begin{align}
	\dd{V}^i + \omega^i{}_jV^j = 0 \,.
\end{align}
Here, the spin-connection 1-form, $\omega$, is antisymmetric and thus, has only one independent component in 2D, $\omega^8{}_9=-\omega^9{}_8$. For a closed curve $\gamma=\partial\mathcal{S}$, the solution to this parallel transport condition is a path-ordered exponential that reduces to a rotation $R(\Phi)\in\text{SO}(2)$ with the angle $\Phi=-\oint_{\gamma}\omega^8{}_9$. Since $\Phi_\text{flat}=2\pi$ for flat polar coordinates, the deficit angle in our case is given by
\begin{align} \label{deltaDef}
	\delta_{\gamma} = \Phi_\text{flat} - \Phi = 2\pi + \oint_{\gamma}\omega^8{}_9 \,.
\end{align}
Indeed, this holonomy is also related to the curvature 2-form, $R^i{}_j=\dd\omega^i{}_j+\omega^i{}_k\wedge\omega^k{}_j$, through $R^8{}_9=\dd{\omega}^8{}_9$ along with Stokes' theorem $\oint_{\partial\mathcal{S}}\omega^8{}_9=\int_\mathcal{S}\dd{\omega}^8{}_9$. However, we employ the boundary integral here to avoid integrating over the unknown physics near the origin.

The transverse part of the metric \eqref{metricAnsatz} in polar coordinates is $\dd{s}_\perp^2 =e^{h(r,\theta)}(\dd{r}^2+r^2\dd{\theta}^2)$, which suggests the orthonormal coframe $e^8=e^{h/2}\dd{r}$ and $e^9=e^{h/2}r\dd{\theta}$. Using Cartan's structure equations, we obtain the singular spin connection component
\begin{align}
	\omega^8{}_9 = \frac{1}{2r}\partial_\theta h\dd{r} - \Big(1 + \frac{r}{2}\partial_r h\Big)\dd{\theta} \,.
\end{align}
Finally, after transforming \eqref{hSolfixed} to polar coordinates, we can compute the integral in \eqref{deltaDef} over a circle with radius $R\gg r_0$ to find
\begin{align}
	\delta_R = 2\pi\bigg(\mu + \frac{2q(2-\mu)}{R} + \cdots\bigg) \,,
\end{align}
which reduces to the expected result as we send $R$ to infinity:
\begin{align}
	\lim_{R\to\infty} \delta_R = 2\pi\mu \,.
\end{align}

Given the asymptotically conical nature of our solution, we should expect our background solution to asymptote to flat spacetime with vanishing gravitational energy since there is no localized energy density at infinity. Such behavior may be quantified in terms of the Brown-York quasi-local energy \cite{Brown:1992br},
\begin{align}
    E_{BY}(R) = \frac{1}{2\pi}\int_{\partial\mathcal{S}}(K - K_0) \,,
\end{align}
where 
\begin{align}
    K = \nabla_i n^i = \frac{1}{\sqrt{g_\perp}}\partial_i\big(\sqrt{g_\perp}e^{-h/2}\delta^i_r\big))
\end{align}
is the extrinsic curvature of the transverse metric in the unit normal radial direction, $K_0$ is the analogous quantity corresponding to flat two-dimensional Euclidean space, and the integral is performed over the boundary of the transverse space at fixed $R \gg r_0$. In terms of our explicit background solutions, we find
\begin{align}
    E_{BY}(R) = -R^{-1} + (1-\mu)R^{-1+\mu} + 2q(2-\mu)^2R^{-2+\mu} + \cdots \,. 
\end{align}
Requiring that the $R\to\infty$ limit of this expression, i.e., the ADM mass, vanishes, thus implies the bound
\begin{align} \label{mubound}
	\mu = \frac{\delta}{2\pi} < 1 \,.
\end{align}
This same bound may also be derived by requiring asymptotic flatness of the full 10D metric\footnote{See also \cite{Sen:2025iuf}.}, $\lim_{r\to\infty}R_\mu{}^\nu = 0$, where the $i \neq j$ components take the form:
\begin{align}
    R_i{}^j = -iq(\mu - 2)r^{-2(1-\mu)}e^{i\theta} + \cdots + \text{h.c.} \,.
\end{align}
We also note that the requirement \eqref{mubound} compliments the general bound of $0\leq\delta<4\pi$ that arises from the ``Elementary Constituents Conjecture'' proposed in \cite{Nevoa:2025xiq}.

Finally, it is also worth pointing out that the Brown-York quasi-local energy serves to justify our previous choice to interpret the complete series solution for the warp factor $f$ in terms of a pure dipole, as discussed after \eqref{Fseries}. If we were to include higher terms in the sum \eqref{Fseries}, their leading arbitrary constants end up being identified with higher multipole coefficients that lead to a quasi-local energy that splits cleanly into the schematic form
\begin{align}
    E_{BY}(R) = &-(\text{flat space})R^{-1} + (\text{monopole})R^{-1+\mu} \nonumber\\
    &+ (\text{dipole})R^{-2+\mu} + (\text{quadrupole})R^{-3+\mu} + \cdots \,.
\end{align}
Put simply, if one were to consider a more complicated solution for $f$ with higher multipoles, those higher moments only appear in terms at increasingly less-relevant order and do not mix into the coefficients of the leading order terms. The dipole part of $f$ thus captures the leading order part of our solution while the higher multipoles contribute only next-order corrections, as we have previously asserted.

\section{Scattering of Probe Particles} \label{sec:SCATTERING}

With the background field profiles in place, we now turn to some further properties of our Alice-vortex solutions. In particular, we study the scattering of scalar and pseudo-scalar degrees of freedom off our background. To first approximation, i.e., neglecting the back-reaction from perturbing the degrees of freedom localized on the brane, this is fixed by the Green's function in this background. Our aim will be to study this scattering problem for a probe scalar and pseudo-scalar, and then for fluctuations of the composite axio-dilaton.

\subsection{Scalar/Pseudo-Scalar Two-Point Functions} \label{subsec:scalar2pt}

Beginning with the simplest case, we consider the addition of scalar/pseudo-scalar $\varphi(x)$ to the original action \eqref{axiodilaton_action}. A canonically normalized kinetic term leads to the equation of motion
\begin{align}
	\frac{1}{\sqrt{-g}} \partial_\mu \left(\sqrt{-g} g^{\mu \nu} \partial_\nu \varphi \right) = 0 \,,
\end{align}
which may be expanded into
\begin{align}
	e^{-f}\eta^{ab}\partial_b\partial_a \varphi + e^{-h}\delta^{ij}\big(\partial_i\partial_j + 4\partial_i f\partial_j \big)\varphi = 0
\end{align}
after applying the metric ansatz \eqref{metricAnsatz}. This expression can be further simplified into a purely transverse equation in polar coordinates by Fourier expanding in worldvolume momentum modes,\footnote{For ease of exposition we absorb the factors of $(2 \pi)$ which normally appear in the Fourier transform into the definition of $\varphi_p$.} $\varphi(x) = \int \dd[8]{p} e^{i\eta_{ab}p^ax^b}\varphi_p(r,\theta)$, which yields
\begin{align} \label{phiEOM}
	e^{-h}\mathcal{L}\,\varphi_p = 0 \qquad\text{with}\qquad
    \mathcal{L} = \Delta + 4\partial_i f\partial_i - e^{h-f}p^2 \,.
\end{align}

Our goal is to derive the transverse Green's function $G(r,\theta,r',\theta')$ associated with \eqref{phiEOM} which satisfies
\begin{align} \label{LGeq}
	e^{-h}\mathcal{L} \, G(r,\theta,r',\theta') = \delta^{(2)}_g(\vec{r}) \,.
\end{align}
Given the asymptotic nature of our metric and axio-dilaton solutions, obtaining an explicit representation of $G$ is best achieved with Born approximation. This is a rather lengthy calculation that we relegate to Appendix \ref{app:scatter}; here we quote only the end result.

The Fourier expansion of the scalar $\varphi$ two-point function (for $r<r'$ and $\theta<\theta'$) takes the form
\begin{align} \label{GDsum}
	G(r,\theta,r',\theta') = \sum_{k,m\in\mathbb{Z}}\mathcal{A}_{m \to m-k}(r,r')e^{ik\theta'}e^{im(\theta-\theta')} \,,
\end{align}
where the $\mathcal{A}_{m \to m-k}(r,r')$ represent ``multipole transition amplitudes'' that quantify how the $m^{\mathrm{th}}$ Fourier modes of $\varphi$ interact and scatter off the $k$-pole part of the brane. To leading order for $r<r'$, the $k=0$ monopole amplitude and $k=-1$ dipole amplitude are given, respectively, by
\begin{align}
	\mathcal{A}_{m \to m}(r,r') = &-\frac{1}{4\pi|p|}\,e^{-\frac{|p|}{1-\mu}\big(r'^{1-\mu}-r^{1-\mu}\big)}\,(rr')^{-\frac{1-\mu}{2}} + \cdots \label{phi0trans} \\
	\mathcal{A}_{m \to m+1}(r,r') = &-\frac{q}{8\pi^2\mu}\,e^{-\frac{|p|}{1-\mu}\big(r'^{1-\mu}-r^{1-\mu}\big)}\,(rr')^{-\frac{1-\mu}{2}}\big(r'^{-\mu} - r^{-\mu}\big) + \cdots \label{phi1trans} \,.
\end{align}
These functions describe precisely how the brane's monopole and dipole moments affect the propagation of a probe scalar by imparting angular momentum. Crucially, this corresponds to a physical observable -- one could, in principle, measure the ratio of the dipole moment to the brane tension, $|q/\mu|$, by performing a scattering experiment.

It is also important to note the relationship that governs the mixing of angular modes in this expression:
\begin{align} \label{scalarmixing}
	m \leftrightarrow m - k \qquad\text{with}\qquad k,m \in \mathbb{Z} \,.
\end{align}
The fact that integer modes are only allowed to mix with other integer modes comes as a direct result of enforcing periodicity on $\varphi$.

For a pseudo-scalar field, which is inherently anti-periodic, the leading order transition amplitudes take the exact same functional form as displayed in \eqref{phi0trans} and \eqref{phi1trans}, however, there is an important difference between this and periodic scalar case; these amplitudes only mix integer with half-integer angular modes, i.e.\
\begin{align} \label{pseudoscalarmixing}
	m \leftrightarrow m - k \qquad\text{with}\qquad k\in\mathbb{Z} \qquad m\in\mathbb{Z}+\frac12 \,.
\end{align}
Naturally, in contrast to the scalar case, this pseudo-scalar mixing structure appears as a direct result of enforcing anti-periodicity in $2\pi$.

\subsection{Axio-Dilaton Two-Point Functions} \label{subsec:axiodilaton2pt}

We now turn to the composite axio-dilaton, which represents a more complicated case due to its non-canonical kinetic term. Instead of a free scalar with a vanishing background, we are now interested in the scattering of perturbations $\delta\tau$--$\delta\bar{\tau}$ on top of our $\tau$--$\bar{\tau}$ background solution. The relevant EOM governing these fluctuations is found by perturbing the background EOM \eqref{EOMtauFull} with $\tau\to\tau+\delta\tau$ and $\bar{\tau}\to\bar{\tau}+\delta\bar{\tau}$, and keeping terms first order in the perturbations:
\begin{align}
    \frac{1}{\sqrt{-g}} \partial_\mu \left(\sqrt{-g} g^{\mu \nu} \partial_\nu \delta\tau \right) - \frac{4\partial_\mu\tau\partial^\mu\delta\tau}{\tau-\bar{\tau}} + 2\frac{\partial_\mu\tau\partial^\mu\tau}{(\tau-\bar{\tau})^2}\big(\delta\tau - \delta\bar{\tau}\big) = 0 \,.
\end{align}
Naturally, we also have the conjugate equation that contains a kinetic term for $\delta\bar{\tau}$. As in the free scalar calculation, we simplify this expression into a transverse coordinate equation by inserting the metric ansatz \eqref{metricAnsatz} and Fourier expanding our perturbations in worldvolume modes with
\begin{align}
	\delta\tau(x) = \int\dd[8]{p}e^{i\eta_{ab}p^ax^b}\delta\tau_p(r,\theta) \,, \qquad\qquad \delta\bar{\tau}(x) = \int\dd[8]{p}e^{i\eta_{ab}p^ax^b}\delta{\bar{\tau}}_p(r,\theta) \,,
\end{align}
where we note that $\delta\bar{\tau}_p= (\delta\tau_{-p})^*$, as expected for a complex scalar pair in momentum space. With this, we arrive at the EOM
\begin{align}
	e^{-h}\bigg[\bigg(\Delta + 4\bigg(\partial_i f - \frac{\partial_i\tau}{\tau - \bar{\tau}}\bigg)\partial_i - e^{h-f}p^2\bigg)\delta\tau_p + 2\frac{\partial_i\tau\partial_i\tau}{(\tau-\bar{\tau})^2}\big(\delta\tau_p - \delta\bar{\tau}_p\big)\bigg] = 0
\end{align}
and its conjugate.

The fact that $\delta\bar{\tau}_p$ appears in the $\delta\tau_p$ EOM above (and vice versa for the conjugate equation) means that we are unable to describe the two equations with two independent scalar derivative operators akin to the scalar operator we encountered in the previous section. Instead, our system of equations is described by a matrix $\mathcal{L}$,
\begin{align} \label{dtauEOM}
	e^{-h}\mathcal{L}\begin{pmatrix} \delta\tau_p \\ \delta\bar{\tau}_p \end{pmatrix} = 0 \qquad\text{with}\qquad
    \mathcal{L} = \begin{pmatrix}
		\mathcal{L}_{\tau\tau} & \mathcal{L}_{\tau\bar{\tau}} \\
		\mathcal{L}_{\bar{\tau}\tau} & \mathcal{L}_{\bar{\tau}\bar{\tau}}
	\end{pmatrix} \,,
\end{align}
where
\begin{align}
	&\mathcal{L}_{\tau\tau} = \big(\mathcal{L}_{\bar{\tau}\bar{\tau}}\big)^* = \Delta + 4\bigg(\partial_i f - \frac{\partial_i\tau}{\tau - \bar{\tau}}\bigg)\partial_i - e^{h-f}p^2 + 2\frac{\partial_i\tau\partial_i\tau}{(\tau-\bar{\tau})^2} \,, \\
    &\mathcal{L}_{\tau\bar{\tau}} = \big(\mathcal{L}_{\bar{\tau}\tau}\big)^*  = -2\frac{\partial_i\tau\partial_i\tau}{(\tau-\bar{\tau})^2} \,.
\end{align}
This structure implies that the associated Green's function will necessarily appear as a matrix whose off-diagonal components describe mixing between $\delta\tau_p$ and $\delta\bar{\tau}_p$. Concretely, we are interested in solutions to the system of equations
\begin{align} \label{LGmateq}
	e^{-h}\mathcal{L} \, G(r,\theta,r',\theta') = \delta^{(2)}_g(\vec{r}) \qquad\text{with}\qquad
    G = \begin{pmatrix}
		G_{\tau\tau} & G_{\tau\bar{\tau}} \\
		G_{\bar{\tau}\tau} & G_{\bar{\tau}\bar{\tau}}
	\end{pmatrix} \,,
\end{align}
where the delta function on the right-hand side should be understood as a matrix in the appropriate way (see the discussion in Appendix \ref{subapp:axiodilaton}).

As in the previous section, each of the two-point functions above is expressed as a series of radially dependent amplitudes over Fourier modes
\begin{align} \label{GDmatsum}
	G_{ij}(r,\theta,r',\theta') = \sum_{k,m\in\mathbb{Z}/2}\mathcal{A}^{(ij)}_{m \to m-k}(r,r')e^{ik\theta'}e^{im(\theta-\theta')} \,,
\end{align}
which, due to the anti-periodic nature of the source and the complicated mixing between the axion and dilaton through $\delta\tau_p$ and $\delta\bar{\tau}_p$, contains terms that mix all half-integer angular modes via the relationship
\begin{align} \label{taumixing}
	m \leftrightarrow m - k \qquad\text{with}\qquad k,m \in \frac{\mathbb{Z}}{2} \,.
\end{align}
This behavior arises because the $\delta\tau_p$ potential contains periodic ($k\in\mathbb{Z}$) terms that originate from the metric warp factors as well as anti-periodic ($k\in\mathbb{Z}+\frac12$) terms coming from the $\tau$ background that are not present in the potential for the free pseudo-scalar. It is not surprising that this twisting between all possible half-integer modes obscures the monodromy that we expect from the doublet $(\delta\tau_p,\delta\bar{\tau}_p)^T$. The monodromy condition \eqref{tauMonodromy} is after all a statement about analytic continuation between sheets, indeed, we should expect that the Green's function matrix for the axio-dilaton should satisfy
\begin{align}
    G(x_{2\pi},x') = \begin{pmatrix} 0 & -1 \\ -1 & 0 \end{pmatrix}G(x,x') \,,
\end{align}
where $x_{2\pi}$ denotes a winding around the defect. In our large-$r$/Born approximation treatment, we work on only the single branch $\theta\in(0,2\pi)$ and consider only perturbative $\tau$-$\bar{\tau}$ mixing, so the exact sheet changing relation is not enforced at any finite order. In the one-sheet expansion, its only remnant is the appearance of half-integer mixing between angular modes.

For explicit functional forms of the partial wave amplitudes, we once again quote the leading result here and refer the reader to Appendix \ref{app:scatter} for higher order expressions and details of their derivation. The dominant channels in the matrix Green's function lie on the diagonals and, for the first few partial wave components, are given by (for $r < r'$)
\begin{align}
	\mathcal{A}^{(\tau\tau)}_{m \to m}(r,r') = &-\frac{1}{4\pi|p|}\,e^{-\frac{|p|}{1-\mu}\big(r'^{1-\mu}-r^{1-\mu}\big)}\,(rr')^{-\frac{1-\mu}{2}} + \cdots \,, \label{tt0trans} \\
	\mathcal{A}^{(\tau\tau)}_{m \to m+\frac12}(r,r') = &\,\frac{\sqrt{q\,(2-\mu)}}{\sqrt{2}\pi|p|}\,e^{-\frac{|p|}{1-\mu}\big(r'^{1-\mu}-r^{1-\mu}\big)}\,(rr')^{-\frac{1-\mu}{2}}r'^{-1/2} + \cdots \,, \label{tt12trans} \\
	\mathcal{A}^{(\tau\tau)}_{m \to m+1}(r,r') = &-\frac{q}{8\pi^2\mu}\,e^{-\frac{|p|}{1-\mu}\big(r'^{1-\mu}-r^{1-\mu}\big)}\,(rr')^{-\frac{1-\mu}{2}}\big(r'^{-\mu} - r^{-\mu}\big) + \cdots \label{tt1trans} \,.
\end{align}
We note that, to the leading order displayed here, the $k=0,-1$ amplitudes are equal the canonical scalar case shown in the last section, though differences do start to appear at next-to-leading order. The off-diagonal amplitudes that correspond to $\tau$-$\bar{\tau}$ mixing, which are generally suppressed by an overall factor of $r^{-2(1-\mu)}$ or more compared to the diagonal components, have the following forms at first order (again, for $r < r'$):
\begin{align}
	\mathcal{A}^{(\tau\bar{\tau})}_{m \to m}(r,r') = &\,\,\frac{\sqrt{q\bar{q}}}{4\pi^2p^2}e^{-\frac{|p|}{1-\mu}(r'^{1-\mu}-r^{1-\mu})}(rr')^{-\frac{1-\mu}{2}}\big(r'^{-2+\mu} - r^{-2+\mu}\big) + \cdots \,, \label{ttb0trans} \\
	\mathcal{A}^{(\tau\bar{\tau})}_{m \to m+\frac12}(r,r') = &\,\,\frac{q\sqrt{\bar{q}\,(2-\mu)}}{\sqrt{2}\pi^2p^2}e^{-\frac{|p|}{1-\mu}\big(r'^{1-\mu}-r^{1-\mu}\big)}(rr')^{-\frac{1-\mu}{2}}r'^{-1/2}r^{-2+\mu} + \cdots \,, \label{ttb12trans} \\
    \mathcal{A}^{(\tau\bar{\tau})}_{m \to m+1}(r,r') = &-\frac{q^{3/2}\sqrt{\bar{q}}(2-\mu)(ic_2 + 4\mu - 8)}{\pi^2p^2(2\mu - 5)}e^{-\frac{|p|}{1-\mu}\big(r'^{1-\mu}-r^{1-\mu}\big)} \nonumber\\
    &\qquad \times (rr')^{-\frac{1-\mu}{2}}r'^{-1/2}r^{-\frac{5-2\mu}{2}} + \cdots \,. \label{ttb1trans}
\end{align}
Expressed as a mode expansion, the complete Green's function for $\tau$ scattering is completely characterized by the relationship \eqref{taumixing} and these radial coefficient functions. Given the fixed nature of the angular dependence, all of the relevant physical information that could be measured by performing a scattering experiment is encoded in these radial amplitudes.

\section{R7-Brane Worldvolume} \label{sec:WORLDVOLUME}

We now consolidate these results to extract some general lessons for the worldvolume theory of R7-branes. To begin, let us recall some earlier lessons on the structure of the worldvolume theory.\footnote{See also \cite{Dierigl:2022reg, Heckman:2025wqd}.} Recall that the supersymmetric $\mathrm{SO}(8)$ 7-brane configuration of F-theory can be realized via a bound state of an $F_L$ and $\Omega$ R7-brane \cite{Dierigl:2022reg} since the net $\mathrm{SL}(2,\mathbb{Z})$ monodromy is given by $\mathrm{diag}(-1,-1)$. Now, as it is a supersymmetric QFT in 8D, this $\mathrm{SO}(8)$ theory eventually flows to an IR trivial fixed point.\footnote{Indeed, the maximum dimension for a superconformal field theory is six \cite{Nahm:1977tg}.} It is natural to ask whether the constituent R7-branes exhibit similar features, or instead have more complicated dynamics.

To begin, we observe that the metric for the $\mathrm{SO}(8)$ stack is not simply given by multiplying the warp factors for the individual R7-branes. Indeed, since it preserves supersymmetry and can be formulated as a local patch of a K3 surface in F-theory, it takes the generic (local) form:
\begin{equation}
ds^2_{\mathrm{SO}(8)} = \eta_{ab} \dd{x}^{a} \dd{x}^{b} + e^{h_{\mathrm{SO}(8)}(z,\overline{z})} \dd{z} \dd{\overline{z}}.
\end{equation}
In particular, there is no warp factor in the worldvolume directions. This is due to the $\Omega$ and $F_L$ R7-branes forming a non-trivial bound state \cite{Dierigl:2022reg}. Indeed, if we compare the structure of the $e^h$ warp factors for the R7-branes and the $\mathrm{SO}(8)$ stack, they do not form a marginal bound state. One consequence of this is that the conical deficit angle for the R7-branes cannot simply be half of that of the $\mathrm{SO}(8)$ stack. In particular, since four $\mathrm{SO}(8)$ stacks build a compact K3 surface in F-theory \cite{Vafa:1996xn, Sen:1996vd}, each generates a deficit angle of $\pi$, so the corresponding value of $\mu = \delta / 2 \pi$ is:
\begin{align}
\mu_{\mathrm{SO}(8)} &= \frac{1}{2} \\
\mu_{\Omega} + \mu_{F_L} &= \mu_{\mathrm{SO}(8)} - \mu_{\mathrm{BIND}} < \frac{1}{2},
\end{align}
since the binding energy is non-zero.

This is already a good indication that the worldvolume theory of the R7-brane consists of more than just free fields. Indeed, returning to our scattering analysis in section \ref{sec:SCATTERING}, we observed a highly non-trivial dependence on the deficit angle $\delta = 2 \pi \mu$. One expects that only simple rational values of $\mu$ would be compatible with a collection of free fields, and it seems implausible that this will happen in the absence of supersymmetry. Though suggestive, this by itself is not a direct proof since we do not (yet) have a way to extract the tension of the R7-brane directly.

Putting all of these elements together provides intriguing evidence that the R7-brane supports a non-trivial 8D QFT, i.e., it is not simply a collection of free fields. Further support for this picture comes from the fact that our family of gravitational backgrounds generically have an intrinsic (gravitational) dipole moment.\footnote{Even if we tune our family of solutions to eliminate the dipole moment, there is still a higher order multipole moment so the rotational symmetry in directions transverse to the brane is generically broken.} This in turn means that there is an intrinsic scale and ``puffing up'' direction for these brane solutions. This is reminiscent of the T-brane configurations for supersymmetric branes found in \cite{Cecotti:2010bp} (see also \cite{Anderson:2013rka, Anderson:2017rpr}), which can be interpreted as non-commutative structure in matrix valued fields of the brane worldvolume. Additionally, the appearance of such an intrinsic scale would seem to suggest that the worldvolume (at least in these backgrounds) has an operator with a non-trivial vacuum expectation value: $\langle \mathcal{O}_{\mathrm{dipole}} \rangle \neq 0$ in order to source the dipole of the bulk solution. In this case, the symmetry being broken can be interpreted as the $U(1)_{\bot}$ isometries transverse to the brane, which in turn suggests the appearance of a massless Goldstone mode.\footnote{The spontaneous symmetry breaking can naturally arise from a radiatively generated potential for a scalar, as in \cite{Coleman:1973jx}.}

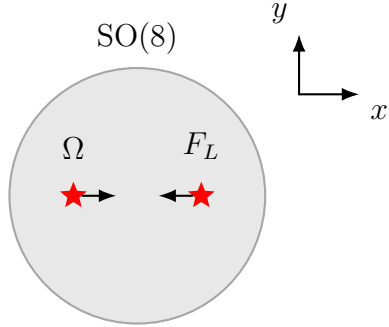
\begin{figure}[t!]
  \centering
  \begin{tikzpicture}[scale=1.0, line cap=round, line join=round, >=Latex]
  \def\R{1.7}
  \def\dx{0.1}
  \path[use as bounding box] ({-\R-\dx},{-\R}) rectangle ({\R-\dx},{\R+0.6});
  \def\ay{0.00}

  \fill[gray!18] (0,0) circle (\R);
  \draw[gray!70, thick] (0,0) circle (\R);

  \node at (0,\R+0.40) {$\mathrm{SO}(8)$};

  \node[star, star points=5, star point ratio=2.25, draw=none, fill=red, minimum size=3.75mm, inner sep=0pt] (L) at (-0.85,\ay) {};
  \node[star, star points=5, star point ratio=2.25, draw=none, fill=red, minimum size=3.75mm, inner sep=0pt] (R) at (0.85,\ay) {};

  \draw[thick,->] (L) -- (-0.28,\ay);
  \draw[thick,->] (R) -- ( 0.28,\ay);

  \node[above=3.5mm] at (L) {$\Omega$};
  \node[above=3.5mm] at (R) {$F_L$};

  \begin{scope}[shift={(2.15,1.35)},overlay]
    \draw[thick,->] (0,0) -- (0.8,0) node[below right] {$x$};
    \draw[thick,->] (0,0) -- (0,0.8) node[above left] {$y$};
  \end{scope}
\end{tikzpicture}
  \caption{Depiction of an $\mathrm{SO}(8)$ 7-brane viewed as a bound state of a $F_L$ and $\Omega$ R7-brane. These R7-branes each come with a dipole moment (black arrows) which must be anti-aligned since there is no net dipole for the $\mathrm{SO}(8)$ 7-brane.}
  \label{fig:dipole_figure}
\end{figure}

Observe also that this dipole interacts non-trivially with other branes. For example, consider a $F_L$ and $\Omega$ R7-brane. The dipoles of these two branes must be exactly anti-aligned simply because the $\mathrm{SO}(8)$ 7-brane does not carry a dipole. This in turn means that there must be \textit{some} interaction between the dipole moments taking place to force this alignment in the first place. See figure \ref{fig:dipole_figure} for a depiction.

Putting all of this together strongly suggests that the 8D worldvolume QFT on an R7-brane
is not simply a collection of free fields. That being said, we leave a more complete treatment of this system for future work.\footnote{It is tempting to speculate that if this theory contains gapless degrees of freedom but is not completely free, that it specifies an 8D non-supersymmetric conformal field theory (CFT). Note, however, that the appearance of a non-zero dipole moment introduces a mass scale so any putative 8D CFT could only arise below this scale in the deep IR anyway. While it could be that only the Goldstone mode associated with breaking the $U(1)_{\bot}$ isometry survives, we find this rather implausible given that the bound state of the $F_L$ and $\Omega$ R7-branes has a far larger number of degrees of freedom.}

\section{Conclusions} \label{sec:CONC}

In this work, we have derived a family of explicit gravitational solutions for Alice-vortices in axio-dilaton gravity. In particular, we determined the far-field profile for the R7-brane of type IIB supergravity in a non-trivial axion background.
After solving the gravitational and axio-dilaton equations of motion subject to the anti-periodic condition $\tau(r,\theta + 2 \pi) = -\bar{\tau}(r, \theta)$, we were able to identify the free parameters in the general solutions with physical data -- multipole moments of the R7-brane. Given the new free parameters that appear at each order in our power series solutions, we restricted to the most generic possible energy-momentum tensor and considered longitudinal components $T_{ab}$ that only contain monopole source terms. Though higher multipoles in the longitudinal energy-momentum tensor components do not affect the qualitative nature of our solutions, we found that the brane \textit{must} carry non-zero higher multipole moments in the transverse energy-momentum tensor components if the $\tau$ profile satisfies $\tau(r,\theta + 2 \pi) = -\bar{\tau}(r, \theta)$. All of these non-zero multipoles associated with the R7-brane have important physical interpretations. We found the expected result that the monopole/brane tension sources a conical deficit angle for vectors parallel transported around the brane, and we argued that this angle should lie in the range $0<\delta<2\pi$ in order to maintain an asymptotically flat spacetime. We also calculated the two-point function of a probe scalar incident on the brane, the functional form of which indicates how different angular momentum modes among the scalar and the brane sources mix and confirms how the multipole moments of the brane show up in asymptotic scattering observables. In the remainder of this section we discuss some avenues for future investigation.

We studied in some detail the behavior of probe scalars propagating on our background solution in Section \ref{sec:SCATTERING} and saw that, subject to the appropriate boundary conditions, they exhibit purely decaying behavior as our solutions asymptote to locally flat space. We have not yet however considered the propagation of more general perturbations on our background, particularly on the gravitational side. It would be interesting to perform a complete study of the stability of gravitational perturbations in our setup; however, the results may already be anticipated to some degree. Given the 8D Poincar\'e invariance along the worldvolume, gravitons naturally separate into distinct transverse-traceless tensor, vector, and scalar sectors. While the vector sector is pure gauge, the effective operator governing dynamics on the 2D transverse space in the tensor sector is necessarily of Sturm-Liouville type, and we can thus anticipate that it will reduce to the same kind of modified Bessel operator that we encountered in Section \ref{sec:SCATTERING}. The worldvolume momentum term that appears in this operator (see \eqref{phiEOM}) provides a positive mass gap and should forbid the presence of normalizable negative modes. The scalar sector will require specific care, however -- there will exist non-trivial mixing when breathing modes (perturbations of the transverse warp factor) couple to axio-dilaton fluctuations through the full Hessian. We also expect that in the case of multiple R7-branes, there will be an unstable mode causing the solution to clump, since it is a non-BPS configuration. A complete stability analysis of the scalar sector is thus certainly warranted in the future.

Establishing the existence of self-consistent gravitational backgrounds provides significant supporting evidence for the existence of R7-branes.
Moreover, the appearance of a whole family of solutions controlled by non-trivial multipole moments provides evidence for a rich class of non-supersymmetric bound states. It would be interesting to extract further properties of such objects.

Our analysis has been at the level of supergravity, and as such, we cannot fully determine the tension of this brane solution. It is natural to expect that a generalization of the holographic analysis presented in \cite{Heckman:2025isn} (see also \cite{Heckman:2024oot, Cummings:2026giw}) can be used to determine this data. The primary complication in this setting is the significant backreaction expected for the gravity dual of the topological symmetry operator of charge conjugation, but we expect the considerations presented here to extend to asymptotically AdS backgrounds.

Indeed, while we have primarily focused on asymptotically flat spacetimes, it is natural to expect our considerations to generalize to gravitational backgrounds of the form $\mathrm{AdS} \times X$. This would be a natural starting point for generating a rather novel class of non-supersymmetric flux compactifications.

We have also provided further evidence for rich worldvolume dynamics of the R7-brane, i.e., it does not appear to simply be a set of free fields. The appearance of an intrinsic dipole scale introduces a length scale into this system, but this does not fully address what happens in the deep IR of this 8D non-supersymmetric QFT. It would be very interesting to study this worldvolume theory further.

\newpage

\section*{Acknowledgments}

We thank O. Bergman, V. Chakrabhavi, M. Dierigl, J. McNamara, M. Montero, and E. Torres for helpful discussions.
AC and JJH thank the 22nd Simons Summer Workshop at the Simons Center for Geometry and Physics
for hospitality during part of this work. The work of AC is supported by the Turkish Fulbright Commission's PhD Grant.
The work of MC, JJH, and CM  is supported by DOE (HEP) Award DE-SC0013528. The work of MC is also supported by
the Slovenian Research Agency (ARRS No. P1-0306) and Fay R. and Eugene L. Langberg Endowed Chair funds.
The work of JJH is also supported by BSF grant 2022100 and a University Research Foundation grant at the
University of Pennsylvania. The work of JK is supported by the the Alexander von Humboldt Foundation.

\appendix

\section{Axio-Dilaton Gravity in $D$-Dimensions} \label{app:DDIM}

In this Appendix, we derive the Einstein's equation given in \eqref{ijEOM}--\eqref{abEOM2} and present their solutions to higher order. To keep the discussion general, instead of restricting ourselves to 10D, we will work with the $D$-dimensional axio-dilaton gravity system described by the action
\begin{equation} \label{axiodilaton_action_inDd}
    S = \frac{1}{2\kappa^2}\int \dd[D]{x} \sqrt{-g} \left(R - \frac{\partial_\mu \tau \partial^\mu \bar{\tau}}{2 (\Im\tau)^2} \right) \,.
\end{equation}

\subsection{Equations of Motion}

We use the same ansatz for the metric as in 10D, which is given by
\begin{equation}
    \dd{s}^2 = e^{f(x^i)}\eta_{ab}\dd{x}^a\dd{x}^b + e^{h(x^i)} \delta_{ij}\dd{x}^i\dd{x}^j \, ,
\end{equation}
where early-alphabet Latin indices $a$, $b$, ...\ refer to the $(D-2)$-dimensional external space with the Minkowski metric $\eta_{ab}$ and mid-alphabet Latin indices $i$, $j$, ...\ refer to the 2-dimensional external space with the Euclidean metric $\delta_{ab}$. With this ansatz in $D$ dimensions, the equation of motion for the axio-dilaton $\tau$ in Eq.~\eqref{EOMtauFull} becomes
\begin{equation} \label{EOMtau_Ddim}
    \partial_i \partial_i \tau + \frac{D-2}{2} \partial_i f \partial_i \tau + i \frac{\partial_i \tau \partial_i \tau}{\Im \tau} = 0 \, .
\end{equation}

The next step is to determine the Einstein equations $G_{\mu \nu} = T_{\mu \nu}$. The non-zero components of the Ricci tensor $R_{\mu \nu}$ and the Ricci scalar $R$ are
\begin{align}
    R_{ab} &= - \frac{1}{2} e^{f-h} \eta_{ab} \left( \partial_i \partial_i f + \frac{D-2}{2} \partial_i f \partial_i f \right) \, , \\
    R_{ij} &= - \frac{D-2}{2} \partial_i \partial_j f - \frac{D-2}{4} \partial_i f \partial_j f - \frac{1}{2} \delta_{ij} \partial_k \partial_k h + \frac{D-2}{4} \left(\partial_i f \partial_j h + \partial_i h \partial_j f - \delta_{ij} \partial_k f \partial_k h \right) \, , \\
    R &= -e^{-h} \left( \left(D-2\right) \partial_i \partial_i f + \frac{1}{4} \left(D-1\right) \left(D-2\right) \partial_i f \partial_i f + \partial_i \partial_i h \right) \, ,
\end{align}
and the longitudinal and transverse components of the Einstein tensor $G_{\mu \nu} = R_{\mu \nu} - \frac{1}{2} g_{\mu \nu} R$ are then
\begin{align}
    G_{ab} &= \frac{1}{2} \eta_{ab} e^{f-h} \left( \left(D-3\right) \partial_i \partial_i f + \frac14 \left(D-2\right) \left(D-3\right) \partial_i f \partial_i f + \partial_i \partial_i h \right) \, , \\
    G_{ij} &= \frac{D-2}{4} \bigg( - 2 \partial_i \partial_j f - \partial_i f \partial_j f + \delta_{ij} \left( 2 \partial_k \partial_k f + \frac{D-1}{2} \partial_k f \partial_k f \right)  \nonumber \\ & \qquad+ \left(\partial_i f \partial_j h + \partial_i h \partial_j f - \delta_{ij} \partial_k f \partial_k h\right) \bigg) \, .
\end{align}

The bulk energy-momentum tensor $T_{\mu \nu}^{(\text{bulk},\tau)}$,
\begin{equation}
    T_{\mu \nu}^{(\text{bulk},\tau)} = \frac{1}{4(\Im\tau)^2}\left(\partial_\mu \tau \partial_\nu \bar \tau + \partial_\nu \tau \partial_\mu \bar \tau - g_{\mu \nu}\partial_\rho\tau\partial^\rho\bar{\tau} \right) \, ,
\end{equation}
is the same in $D$-dimensions as in 10D, and has the longitudinal and transverse components
\begin{align}
    T_{ab}^{(\text{bulk},\tau)} &= -\eta_{ab} e^{f-h} \frac{\partial_i \tau \partial_i \bar{\tau}}{4 (\Im \tau)^2} \, ,\\
    T_{ij}^{(\text{bulk},\tau)} &= \frac{1}{4 (\Im \tau)^2} \left( \partial_i \tau \partial_j \bar{\tau} + \partial_j \tau \partial_i \bar{\tau} - \delta_{ij} \partial_k \tau \partial_k \bar{\tau} \right) \, .
\end{align}

The first equation to consider is $G_{ii} + G_{jj} = T_{ii} + T_{jj}$ ($i \neq j$). Since $T_{ii} + T_{jj} = 0$, this equation results in a constraint on $f$ independent of $h$ or $\tau$:
\begin{equation} \label{fcon_Ddim}
    \frac{D-2}{2} \left( \partial_i \partial_i f + \frac{D-2}{2} \partial_i f \partial_i f \right) = 0 \, ,
\end{equation}
which turns into \eqref{fcon} for $D=10$. This equation can be converted to a Laplace equation by the substitution $f = \frac{2}{D-2} \log F$:
\begin{equation}
    f = \frac{2}{D-2} \log F \quad \Rightarrow \quad \frac{\partial_i \partial_i F}{F} = 0 \, .
\end{equation}

There are three more independent Einstein equations. After switching to holomorphic coordinates, the $G_{zz} = T_{zz}$ and $G_{\bar{z} \bar{z}} = T_{\bar{z} \bar{z}}$ equations yield \eqref{abEOM1} and \eqref{abEOM2}:
\begin{align}
    -\frac{D-2}{2}\left(\partial^2 f + \frac12(\partial f)^2 - \partial f \partial h\right) &= \frac{\partial \tau \partial \bar{\tau}}{2(\Im\tau)^2} \, , \label{ijEq1_Ddim}\\
    -\frac{D-2}{2}\left(\bar{\partial}^2 f + \frac12(\bar{\partial} f)^2 - \bar{\partial}f\bar{\partial}h\right) &= \frac{\bar{\partial} \tau \bar{\partial} \bar{\tau}}{2(\Im\tau)^2} \,. \label{ijEq2_Ddim}
\end{align}
Moreover, matching the longitudinal components, $G_{ab} = T_{ab}$, gives
\begin{equation}
    \left(D-3\right) \left(\partial_i \partial_i f + \frac{D-2}{4} \partial_i f \partial_i f\right) + \partial_i \partial_i h = - \frac{\partial_i \tau \partial_i \bar{\tau}}{2 \left(\Im \tau\right)^2} \, . \label{abEq_Ddim}
\end{equation}
This turns into \eqref{ijEOM} after substituting \eqref{fcon_Ddim} and switching to holomorphic coordinates.\footnote{However, it should be noted that \eqref{abEq_Ddim} is not independent, because it can be derived from (\ref{ijEq1_Ddim}--\ref{ijEq2_Ddim}) and the equation of motion for $\tau$ in \eqref{EOMtau_Ddim}. This is done by first redefining
\begin{equation}
    h(z,\bar{z}) = h_0(z,\bar{z}) + \frac{D-1}{2} f(z,\bar{z}) + \log\big(\partial f \,\bar{\partial} f\big) \, ,
\end{equation}
using which one can rewrite Eqs.~(\ref{ijEq1_Ddim}-\ref{ijEq2_Ddim}) as
\begin{equation} \label{h0Eq_Ddim}
    \partial f \partial h_0 = \frac{\partial \tau \partial \bar{\tau}}{(D-2) (\Im \tau)^2} \, , \qquad \bar{\partial} f \bar{\partial} h_0 = \frac{\bar{\partial} \tau \bar{\partial} \bar{\tau}}{(D-2) (\Im \tau)^2} \, .
\end{equation}
Taking the antiholomorphic derivative of the first of these equations, one obtains
\begin{align}
    \partial \bar{\partial} f \partial h_0 + \partial f \partial \bar{\partial} h_0 &= \frac{1}{D-2} \left[ \frac{\partial \bar{\tau}}{(\Im \tau)^2} \left(\partial \bar{\partial} \tau + i \frac{\partial \tau \bar{\partial} \tau}{\Im \tau} \right) + \frac{\partial \tau}{(\Im \tau)^2} \left( \partial \bar{\partial} \bar{\tau} - i \frac{\partial \bar{\tau} \bar{\partial} \bar{\tau}}{\Im \tau} \right) \right] \\
    &= - \frac{1}{4 (\Im \tau)^2} \Big[ \partial f \left(\partial \tau \bar{\partial} \bar{\tau} + \partial \bar{\tau} \bar{\partial} \tau \right) + 2 \bar{\partial} f \partial \tau \partial \bar{\tau} \Big] \,,
\end{align}
where we substituted the equation of motion for $\tau$ \eqref{EOMtau_Ddim} and its complex conjugate in the second line. Noting that
\begin{equation}
    \partial \bar{\partial} f \partial h_0 = - \frac{D-2}{2} \bar{\partial} f \partial f \partial h_0 = - \frac{\bar{\partial} f \partial \tau \partial \bar{\tau}}{2 (\Im \tau)^2} \, ,
\end{equation}
where we used \eqref{fcon_Ddim} and \eqref{h0Eq_Ddim}, one obtains
\begin{align}
	\partial f\left(\partial\bar{\partial} h_0 + \frac{\partial\tau\bar{\partial}\bar{\tau} + \bar{\partial}\tau\partial\bar{\tau}}{4(\Im\tau)^2}\right) = 0 \,,
\end{align}
which leads to Eq.~\eqref{abEq_Ddim} as long as $f$ is nontrivial and real.}

If we also include the singular contribution to the energy-momentum tensor sourced by the brane by writing the total energy-momentum tensor as $T_{\mu \nu} = T_{\mu \nu}^{(\text{bulk},\tau)} + T_{\mu \nu}^{\text{(brane)}}$ with the components of $T_{\mu \nu}^{\text{(brane)}}$ given in \eqref{EMsources}, we only need to modify the Einstein's equations $G_{ab} = T_{ab}$ and $G_{ii} + G_{jj} = T_{ii} + T_{jj}$ ($i \neq j$)\footnote{In holomorphic coordinates, the latter corresponds to the equation $G_{z \bar{z}} = T_{z \bar{z}}$.}, which can be written respectively as
\begin{align}
    e^{-h}\Bigg(\left(D-3\right) \left(\partial_i \partial_i f + \frac{D-2}{4} \partial_i f \partial_i f\right) &+ \partial_i \partial_i h + \frac{\partial_i \tau \partial_i \bar{\tau}}{2 \left(\Im \tau\right)^2}\Bigg)  \nonumber \\ &= - 4\pi \Big(\mu\,\delta^{(2)}_g(\vec{r}) -  Q^i\nabla_i\delta^{(2)}_g(\vec{r}) + \cdots\Big) \, , \\
    e^{-h} \frac{D-2}{2}\left( \partial_i \partial_i f + \frac{D-2}{2} \partial_i f \partial_i f \right) &= 4\pi \Big(q^i\nabla_i\delta^{(2)}_g(\vec{r}) + \cdots\Big) \,,
\end{align}
which reproduces \eqref{aaEq}--\eqref{iiEq}. Here, $\delta^{(2)}_g(\vec{r})$ is the 2D curved-space delta function, which includes an extra factor of $(g_{\perp})^{-1/2} = e^{-h}$ that cancels the $e^{-h}$ factor appearing on the left-hand side of these equations.

\subsection{Asymptotic Solutions}

The solutions to the equations of motion in $D$-dimensions proceed almost the same way as in 10D. The equations have the same conformal symmetry $z \to z^\prime(z)$. Therefore, one can first find a solution with (here we set $a \equiv 1$):
\begin{equation}
    f = \frac{2}{D-2} \log \left( 1+ \frac{1}{z} + \frac{1}{\bar{z}} \right) \, ,
\end{equation}
and then obtain a more general solution by a holomorphic transformation $z \to z^\prime(z)$.

To find a solution, we will write $h$ and $\tau$ as power series in $z^{-1/2}$ and $\overline{z}^{-1/2}$:
\begin{align}
    h(z,\bar{z}) &= -\frac{D-3}{2} f(z,\bar{z}) - b \log(z \bar{z}) + \sum_{m,n=0}^{\infty} \beta_{m,n} z^{-m/2} \bar{z}^{-n/2} \, , \\
    \tau(z,\bar{z}) &= i \lambda \sum_{m,n = 0}^{\infty} i^{m+n} \gamma_{m,n} z^{-m/2} \bar{z}^{-n/2} \,.
\end{align}
The motivation to include the term proportional to $f$ in the expansion for $h$ is to make the coefficients $\beta_{m,n}$ and $\gamma_{m,n}$ independent of $D$, as $D$ drops out of the equations after substituting these expressions for $f$, $h$, and $\tau$. The factors of $i$ in the expansion of $\tau$ are included to conveniently make the coefficients $\gamma_{m,n}$ Hermitian, which is different from the convention for $c_{m,n}$ in the main text.

The constraints on the coefficients $\beta_{m,n}$ and $\gamma_{m,n}$ are the same as in the 10D case. The reality and periodicity of $h$ lead to the restrictions
\begin{equation}
    \left((-1)^{m+n} - 1\right) \beta_{m,n} =0  \, , \qquad \ \overline{\beta}_{m,n} = \beta_{n,m} \, ,
\end{equation}
and the monodromy condition $\tau(r,\theta + 2\pi) = - \bar{\tau}(r,\theta)$ leads to the Hermiticity of the coefficients $\overline{\gamma}_{m,n} = \gamma_{n,m}$. We set $\gamma_{0,0} = 1$.

As in the 10D case, the coefficients $\gamma_{i,0}$ remain undetermined. The solutions for the first few coefficients up to order $r^{-5/2}$ are then
{\allowdisplaybreaks\begin{align*}
    \gamma_{1,1} &= \gamma_{0,1}\gamma_{1,0} \, ,\\
    \gamma_{2,1} &= \frac{1}{2} \gamma_{0,1} \left(\gamma_{1,0}^2+2 \gamma_{2,0}+1\right) \,, \\
    \gamma_{3,1} &= \frac{1}{6} \gamma_{0,1} \left(\gamma_{1,0}^3+4 \gamma_{2,0} \gamma_{1,0}+3 \gamma_{1,0}+6 \gamma_{3,0}\right) \,,\\
    \gamma_{4,1} &= \frac{1}{24} \gamma_{0,1} \left(\gamma_{1,0}^4+4 \gamma_{2,0} \gamma_{1,0}^2+6 \gamma_{1,0}^2+24 \gamma_{3,0} \gamma_{1,0}+12 \gamma_{2,0}+24 \gamma_{4,0}+9\right) \,,\\
    \gamma_{2,2} &= \frac{1}{4} \left(2 \gamma_{1,0}^2 \gamma_{0,1}^2+2 \gamma_{2,0} \gamma_{0,1}^2+\gamma_{0,1}^2+2 \gamma_{0,2} \gamma_{1,0}^2+\gamma_{1,0}^2+2 \gamma_{0,2}+4 \gamma_{0,2} \gamma_{2,0}+2 \gamma_{2,0}\right) \,, \\
    \gamma_{3,2} &= \frac{1}{12} \left(4 \gamma_{0,1}^2 \gamma_{1,0}^3+2 \gamma_{0,2} \gamma_{1,0}^3+\gamma_{1,0}^3+9 \gamma_{0,1}^2 \gamma_{1,0}+6 \gamma_{0,2} \gamma_{1,0}+10 \gamma_{0,1}^2 \gamma_{2,0} \gamma_{1,0} \right.\\ & \qquad \left.+8 \gamma_{0,2} \gamma_{2,0} \gamma_{1,0}+6 \gamma_{2,0} \gamma_{1,0}+4 \gamma_{1,0}+6 \gamma_{0,1}^2 \gamma_{3,0}+12 \gamma_{0,2} \gamma_{3,0}+6 \gamma_{3,0}\right) \,,
\end{align*}}
and
{\allowdisplaybreaks\begin{align*}
    \beta_{2,0} &= \frac{1}{8} \left(2 \gamma_{2,0} \gamma_{1,0}^2+\gamma_{1,0}^2-6 \gamma_{3,0} \gamma_{1,0}+4 \gamma_{2,0}^2\right) \, , \\
    \beta_{4,0} &= \frac{1}{16} \left(8 \gamma_{2,0}^3+3 \gamma_{1,0}^2 \gamma_{2,0}^2+4 \gamma_{2,0}^2+2 \gamma_{1,0}^2 \gamma_{2,0}-12 \gamma_{1,0} \gamma_{3,0} \gamma_{2,0} \right.\\ & \qquad \left.-16 \gamma_{4,0} \gamma_{2,0}+9 \gamma_{3,0}^2 -6 \gamma_{1,0} \gamma_{3,0}-2 \gamma_{1,0}^2 \gamma_{4,0}+10 \gamma_{1,0} \gamma_{5,0}\right) \, , \\
    \beta_{1,1} &= -\frac{1}{2} \gamma_{0,1} \gamma_{1,0} \, , \\
    \beta_{3,1} &= \frac{1}{12} \gamma_{0,1} \left(\gamma_{1,0}^3+\left(1-6 \gamma_{2,0}\right) \gamma_{1,0}+6 \gamma_{3,0}\right) \, , \\
    \beta_{2,2} &= \frac{1}{8} \left(\left(-\gamma_{1,0}^2+2 \gamma_{2,0}+1\right) \gamma_{0,1}^2+\left(2 \gamma_{0,2}+1\right) \gamma_{1,0}^2-4 \gamma_{0,2} \gamma_{2,0}\right) \, .
\end{align*}}

Furthermore, one can also express the coefficient $b$ of the logarithmic term in $h$ in terms of $\gamma_{1,0}$ as
\begin{equation}
    b = \frac{1}{8} \left(16 + \gamma_{1,0}^2\right) = \frac{1}{8} \left(16 + \gamma_{0,1}^2 \right) \, ,
\end{equation}
from which one also obtains the restriction that $\gamma_{1,0}^2 = (\overline{\gamma}_{1,0})^2$. Using the condition that $b = \mu <1$, it is possible to write $\gamma_{1,0}$ in terms of $b$ as
\begin{equation}
    \gamma_{1,0} = \pm 2 i \sqrt{2(2-b)} \, ,
\end{equation}
where the two choices for the sign of $\gamma_{1,0}$ correspond to a choice for the sign of the axion $C_0= \Re \tau$. Replacing $\gamma_{1,0}$ and $\gamma_{0,1}$ in the expressions for the coefficients $\gamma_{m,n}$ and $\beta_{m,n}$ and identifying the undetermined coefficients as $c_n = i^{1+n}\gamma_{n,0}$, one obtains \eqref{hSol}-\eqref{ImtauSol} for $D=10$.

\section{Gauss's Law with Multipole Sources} \label{app:GL}

Here we present details of the calculations in Section \ref{sec:OriginSols} that led to the identification of multipole moments in the asymptotic warp factors and $\tau$ solutions, \eqref{aDefs}-\eqref{Qcons}. We are able to glean this information about our solutions in the near-brane region, despite the fact that the solutions are only applicable in the large $r$ limit, by applying the divergence theorem and Gauss's Law.

The first Einstein's equation that we consider, \eqref{iiEq}, takes the form
\begin{align} \label{DfS}
	\Delta f = S \qquad\text{with}\qquad \text{supp}\,S \subset \{r<r_c\} \,,
\end{align}
where $S$ represents a source that is supported only in the brane with core radius $r_c$. Focusing only on the monopole and dipole terms, we assume this source may be approximated by a delta function and its first derivative. Our goal is to establish a realization of Gauss's Law using \eqref{DfS} that allows us to determine the arbitrary constants in the solution for $f$ in \eqref{fSol}. To keep our treatment general, we will take the constant $a$ in \eqref{fSolz} to be a complex number $a = a_x + i a_y$, so \eqref{fSol} gets modified to
\begin{equation}
    f(r,\theta) = \frac{1}{4} \log \left( 1 + \frac{2(a_x \cos\theta + a_y \sin\theta)}{r} \right) \,.
\end{equation}

We work on a disk\footnote{If one wished to be completely agnostic about the source profile on the brane, it would instead be necessary to excise the core completely and work on an annulus $A_{r_0,R}=\{r_0<r<R\}$ with $r_0>r_c$. By assuming a delta function profile and that \eqref{DfS} is satisfied in a distributional sense, we are able to send $r_c,r_0\to0$ and work on a disk.} $D_R=\{r<R\}$ of radius $R>r_c$ that we take to be large enough that our asymptotic series are accurate. We also define the current
\begin{align}
	J^i = \xi\partial^i f - f\partial^i\xi
\end{align}
in terms of a smooth test function that is harmonic on all of $D_R$,
\begin{align} \label{xiDef}
	\xi(r,\theta) = b_0 + r\big(b_x\cos\theta + b_y\sin\theta\big) \,,
\end{align}
where the $b_i$ are arbitrary constants. The current above corresponds to a Gauss-Law flux through the boundary of our disk,
\begin{align} \label{GLflux}
	I(R) = \oint_{\partial D_R}\dd{s}n_iJ^i = \oint_{r=R}\dd{\theta}r\big(\xi\partial_r f - f\partial_r\xi\big) \,,
\end{align}
which we can relate to \eqref{DfS} via the divergence theorem,
\begin{align} \label{divthm}
	I(R) = \int_{D_R}\dd[2]{x}\partial_iJ^i = \int_{D_R}\dd{r}\dd{\theta}r\xi\Delta f = \int_{D_R}\dd{r}\dd{\theta}r\xi S \,,
\end{align}
where we have used that $\Delta\xi=0$.

With the considerations above we are in a position to actually compute the integrals. The boundary integrals in \eqref{GLflux} are valid for any large $R$, so we can insert \eqref{fSol} and \eqref{xiDef} to find
\begin{align}
	\oint_{r=R}\dd{\theta}r\big(\xi\partial_r f - f\partial_r\xi\big) = -\pi\big(a_x b_x + a_y b_y\big) \,.
\end{align}
Finally, the right side of \eqref{divthm} can be evaluated directly using delta functions after noting that \eqref{iiEq} implies the specific source
\begin{align}
	S = \pi q^i\partial_i\delta^{(2)}_g(\vec{r}) \,.
\end{align}
With this, we simply equate the evaluations of \eqref{GLflux} and \eqref{divthm} and collect the result in terms of the $b_i$ to arrive at the generalized version of \eqref{aDefs}
\begin{align}
	a_x = q_x \qquad\qquad a_y = q_y \,.
\end{align}
The derivation of \eqref{bDef} and \eqref{Qcons} then follows in the exact same way as above, with the caveats that $S$ contains a monopole term and that $h$ must be represented by the first few terms in its asymptotic series instead of an exact solution.

\section{More on the Two-Point Functions} \label{app:scatter}

In this Appendix, we provide detailed derivations of the Green's functions presented in Section \ref{sec:SCATTERING}.

\subsection{Scalar/Pseudo-Scalar} \label{subapp:scalar}

Beginning with the canonical scalar case in Section \ref{subsec:scalar2pt}, our goal is to find an explicit asymptotic expression for the Green's function governed by \eqref{LGeq}, repeated here for convenience:
\begin{align}
	e^{-h} \left( \mathcal{L} \, G(x,x') \right) = \delta^{(2)}_g(x-x')
    \qquad\text{with}\qquad
    \mathcal{L} = \Delta + 4\partial_i f\partial_i - e^{h-f}p^2 \,,
\end{align}
where we denote the transverse coordinates collectively as $x$. The derivative cross terms in $\mathcal{L}$ make this a tricky problem to solve directly, so we begin by performing a diagonalizing similarity transformation in terms of $\Phi = 2f$,
\begin{align} \label{tLDef}
	\widetilde{\mathcal{L}} = e^\Phi \mathcal{L} e^{-\Phi} &= \Delta - \Delta\Phi - (\partial_i\Phi)^2 - e^{h-f}p^2 \nonumber\\
    &= \Delta - 4(\partial_i f)^2 - e^{h-f}p^2 \,,
\end{align}
where we have used the equation of motion for $f$ to arrive at the second line. This new derivative operator, $\widetilde{\mathcal{L}}$, is composed only of a Laplacian and background fields, with no derivative cross terms between background fields and the Green's function. Using the definition \eqref{tLDef}, we may now search for solutions to the transformed Green's function equation
\begin{align} \label{tLGeq}
	\widetilde{\mathcal{L}} \, \widetilde{G}(x,x') = \delta^{(2)}(x-x') = \frac{\delta(r-r')\delta(\theta-\theta')}{r} \,,
\end{align}
which is expressed in flat space by recalling that $\sqrt{g_\perp}=re^h$ in polar coordinates. After finding a solution $\widetilde{G}$, we can then relate that result back to the original Green's function of interest via the relation
\begin{align} \label{GtG}
	G(x,x') = e^{-\Phi(x)}e^{\Phi(x')}\widetilde{G}(x,x') \,.
\end{align}

Given the asymptotic nature of our background solutions that enter into the potential portion of $\widetilde{\mathcal{L}}$, it is natural to also calculate an asymptotic representation of our Green's function. We thus expand the complete equation \eqref{tLGeq} as
\begin{align} \label{LtGexp}
	\big(\widetilde{\mathcal{L}}_0 + \delta \widetilde{\mathcal{L}}\big)\big(\widetilde{G}_0 + \widetilde{G}_1 + \cdots\big) = \delta^{(2)}(x-x') \,,
\end{align}
where the zeroth order Green's function $\widetilde{G}_0$ is defined as the integral kernel of the derivative operator $\widetilde{\mathcal{L}}_0$ so that
\begin{align} \label{G0Eq}
    \widetilde{\mathcal{L}}_0 \widetilde{G}_0(x,x^\prime) = \delta^{(2)}(x-x') \,.
\end{align}
This allows us to expand \eqref{LtGexp} and find that the first order Born correction is given by
\begin{align} \label{tG1Int}
	\widetilde{G}_1(x,x') = -\int\dd[2]{s}\widetilde{G}_0(x,s)\delta\widetilde{\mathcal{L}}(s)\widetilde{G}_0(s,x') \,.
\end{align}
Once we have solved for $\widetilde{G}_0$ and $\widetilde{G}_1$, it is then straightforward to convert them back to their original un-tilded versions via \eqref{GtG}.

Moving forward, we use the fact that $\widetilde{G}_0$ may be written as a sum over Fourier modes,
\begin{align} \label{tG0sum}
	\widetilde{G}_0(x,x') = \frac{1}{2\pi}\sum_{m} e^{im(\theta - \theta')}g_m(r,r') \,,
\end{align}
to reduce \eqref{G0Eq} to a purely radial expression. This is achieved by first noting that the $\theta$ portion of the Laplacian in \eqref{tLDef} produces a simple factor of $-m^2/r^2$ when it acts on the $\widetilde{G}_0$ above, so that we can expand the complete derivative operator as
\begin{align}
    \widetilde{\mathcal{L}} &= \widetilde{\mathcal{L}}_0 + \delta\widetilde{\mathcal{L}} \nonumber\\
    &= \pdv[2]{r} + r^{-1}\pdv{r} - m^2r^{-2} - p^2\Big(r^{-2\mu} - \big(4\sqrt{q\bar{q}}(2-\mu) - 2qe^{-i\theta} - 2\bar{q}e^{i\theta}\big)r^{-1-2\mu} + \cdots\Big) \,.
\end{align}
The potential that appears in zeroth order derivative operator is defined by the lowest order term in the expanded potential, which, given the previously established bound of $0<\mu<1$, implies
\begin{align}
    \widetilde{\mathcal{L}}_0 = \pdv[2]{r} + r^{-1}\pdv{r} - p^2r^{-2\mu} \,.
\end{align}
After employing the identity $\delta(\theta-\theta') = \frac{1}{2\pi}\sum e^{im(\theta-\theta')}$, this reduces \eqref{G0Eq} to the $m$-independent radial Green's function equation in terms of $g_m(r,r')=g(r,r')$:
\begin{align} \label{tLgEq}
	\widetilde{\mathcal{L}}_{0}\,g(r,r') = \frac{\delta(r - r')}{r} \,.
\end{align}
The higher order parts of the full potential are then relegated to $\delta\widetilde{\mathcal{L}}$, which is also expressible as a sum over Fourier modes:
\begin{align} \label{dLsum}
	\delta\widetilde{\mathcal{L}}(y) = \sum_{k}e^{ik\theta}u_k(r) \,,
\end{align}
where the first few radial potential functions are identified as
\begin{align} \label{ukDefs}
	&u_0(r) = 4(2-\mu)p^2\sqrt{q\bar{q}}\,r^{-1-2\mu} - m^2r^{-2} + \cdots \,, \\[0.4em]
    &u_{-1}(r) = \big(u_{1}(r)\big)^* = 2p^2q\,r^{-1-2\mu} + \cdots \,.
\end{align}

Before attempting to solve \eqref{tLgEq}, it is also important to note that the considerations above allow us to compute the angular integral\footnote{We note that the radial integral is only taken over the region of validity of our background solutions, $r>r_0$, which encodes our ignorance of the microscopic core physics as an effective boundary condition at $r=r_0$.} in \eqref{tG1Int} after substituting \eqref{tG0sum} and \eqref{dLsum}, leading to
\begin{align} \label{G1int}
	\widetilde{G}_1(x,x') &= -\frac{1}{(2\pi)^2}\int_{r_0}^\infty\dd{r_s}r_s\int_0^{2\pi}\dd{\theta_s}\sum_{k,m,n}g(r,r_s)u_k(r_s)g(r_s,r')e^{i(m\theta - n\theta')}e^{i(-m + k + n)\theta_s} \nonumber\\
	&= -\frac{1}{2\pi}\int_{r_0}^\infty \dd{r_s}r_s\sum_{k,m}g(r,r_s)u_k(r_s)g(r_s,r')e^{i(m\theta - (m-k)\theta')} \,,
\end{align}
where we have used the orthogonality condition
\begin{align}
	\frac{1}{2\pi}\int_0^{2\pi}\dd{\theta}e^{il\theta} = \delta_{l,0} \,,
\end{align}
with $l = -m + k + n$. This delta implies the angular mode mixing relationships described in main text after equation \eqref{scalarmixing} for the scalar and \eqref{pseudoscalarmixing} for the pseudo-scalar after noting that the (anti-)periodicity of our canonical (pseudo-)scalar is enforced on its Green's function via the summation parameter $m$ running over (half-)integers. The potential expansion on the other hand, is given by a sum over integers in both cases, in line with the periodicity of $f$ and $h$.

The radial Green's function $g(r,r')$ is constructed following the standard Sturm-Liouville procedure, which begins by identifying the independent growing and decaying solutions to the homogeneous equation associated with the zeroth order part of \eqref{tLgEq},
\begin{align}
	\widetilde{\mathcal{L}}_0 \psi(r) = \psi''(r) + \frac{1}{r}\psi'(r) - p^2r^{-2\mu}\psi(r) = 0 \,.
\end{align}
This differential equation can be recast as a modified Bessel equation and thus has a closed-form solution\footnote{We also note the absence of arbitrary constant prefactors in this solution; such constants inevitably cancel with the Wronskian in \eqref{radialGreen}.} given by a linear combination of the modified Bessel functions of the first and second kind, order zero:
\begin{align}
	\psi(r) = I_{0}\big(\alpha r^\beta\big) + K_{0}\big(\alpha r^\beta\big) \qquad\text{where}\qquad \alpha = \frac{\abs{p}}{1-\mu}\,, \qquad \beta = 1-\mu \,.
\end{align}
For our purposes, it is sensible to represent the two linearly independent solutions in their asymptotic forms as
\begin{align}
	\psi^\pm(r) = e^{\pm\alpha r^\beta}r^{-\frac{\beta}{2}}\bigg(1 \pm \frac{1}{8\alpha}r^{-1+\mu} + \cdots\bigg) \,.
\end{align}
We may now build $g(r,r')$ from from the basis of solutions $\psi^\pm(r)$ in terms of interior and exterior solutions subject to appropriate boundary conditions:
\begin{align} \label{radialGreen}
	g(r,r') = \begin{cases}
		\frac{\psi^<(r)\psi^>(r')}{W} & r<r' \\
		\frac{\psi^<(r')\psi^>(r)}{W} & r>r'
	\end{cases} \,,
\end{align}
where $W$ is the Wronskian of $\psi^<$ and $\psi^>$. The exterior solution $\psi^>$ should decay as $r\to\infty$, so it is naturally assigned as $\psi^>=\psi^-$. Our ignorance of the microscopic core physics is best captured by a general Robin condition on the interior solution, which may be parameterized by $\psi^<=\psi^+ + \sigma \psi^-$ where $\sigma$ is a constant (essentially a Robin parameter) that encodes imperfect absorption. Moving forward, we will assume that the core is purely absorbing so that $\sigma=0$. With these assignments, the Wronskian is $W=-2|p|$, and we find that the zeroth order Green's function for $r<r'$ and $\theta<\theta'$ is given by
\begin{align} \label{tG0}
	\widetilde{G}_0(x,x') = -\frac{1}{4\pi|p|}\sum_{m} &\,\,e^{im(\theta' - \theta)}\,e^{-\frac{|p|}{1-\mu}(r'^{1-\mu}-r^{1-\mu})} \nonumber\\
    &\times(rr')^{-\frac{1-\mu}{2}}\bigg(1 - \frac{1-\mu}{8|p|}\big(r'^{-1+\mu} - r^{-1+\mu}\big) + \cdots\bigg) \,.
\end{align}
It is also worth noting that if we allow for imperfect absorption, $\sigma \neq0$, then the resulting Green's function will carry additional terms with a factor of $\sigma e^{-2\frac{p}{1-\mu}r^{1-\mu}}$, which leads them to being highly suppressed in the large $r$ limit.

Our next task is to calculate the first Born correction to $\widetilde{G}_0$. The precise expression that then needs to be evaluated is \eqref{G1int}:
\begin{align} \label{tG1Def}
	\widetilde{G}_1(x,x') = -\frac{1}{(2\pi)^2}\sum_{m,k}e^{i(m\theta - (m-k)\theta')}\int_{r_0}^\infty \dd{r_s}r_s\,g(r,r_s)u_k(r_s)g(r_s,r') \,,
\end{align}
which effectively breaks up into three integrals over the regions $[r_0,r]$, $[r,r']$, and $[r',\infty)$. It is possible to obtain rather involved closed form expressions for each of these integrals in terms of incomplete $\Gamma$-functions, though it is more instructive to use their asymptotic approximations. We omit these intermediate results for the sake of brevity, but with them in hand we can finally construct the complete Green's function up to the first level of Born approximation using \eqref{GtG},
\begin{align} \label{GtGexp}
	G(x,x') = e^{-\Phi(x)}e^{\Phi(x')}\big(\widetilde{G}_0(x,x') + \widetilde{G}_1(x,x') + \cdots\big) \,,
\end{align}
where the transformation operator must be expanded using the background $f$ solution as
\begin{align} \label{simtrans}
	e^{\Phi(x)} = 1 - \frac{qe^{-i\theta} + \bar{q}e^{i\theta}}{2r} + \cdots \,.
\end{align}
We also note that, since this transformation back to the original Green's function problem introduces extra angular factors, it is also necessary to shift the sums over $m$ and $k$ in \eqref{tG0} and \eqref{tG1Def} so that modes associated with powers of $e^{i\theta}$ and $e^{i\theta'}$ are consistently identified in the final un-tilded expression of $G$.

Finally, after performing the transformation \eqref{GtGexp} and collecting terms, we are left with \eqref{GDsum}, restated here:
\begin{align}
	G(x,x') = \sum_{k,m}\mathcal{A}_{m \to m-k}(r,r')e^{ik\theta'}e^{im(\theta-\theta')} \,,
\end{align}
where the explicit NLO $k=0$ monopole and $k=-1$ dipole transition amplitudes are given by
\begin{align}
	\mathcal{A}_{m \to m}(r,r') = &-\frac{1}{4\pi}\,e^{-\frac{|p|}{1-\mu}(r'^{1-\mu}-r^{1-\mu})}(rr')^{-\frac{1-\mu}{2}} \nonumber\\
    & \qquad \times\bigg[\frac{1}{|p|} - \frac{m^2}{4\pi p^2(1-\mu)}(rr')^{-1}\big(r'^{\mu} - r^{\mu}\big) \nonumber\\
    & \qquad\qquad + \frac{\sqrt{q\bar{q}}\,(2-\mu)}{\pi\mu}(rr')^{-\mu}\big(r'^\mu - r^\mu\big) + \cdots\bigg] \,, \\
	\mathcal{A}_{m \to m+1}(r,r') = &-\frac{q}{8\pi^2}\,e^{-\frac{|p|}{1-\mu}(r'^{1-\mu}-r^{1-\mu})}(rr')^{-\frac{1-\mu}{2}} \nonumber\\
    & \qquad \times\bigg[\frac{1}{\mu}\big(r'^{-\mu} - r^{-\mu}\big) - \frac{m^2}{4p^2(1-\mu)}r^{-1}r'^{-2}\big(rr'^{\mu} - r'r^{\mu}\big) \nonumber\\
    & \qquad\qquad + \frac{\sqrt{q\bar{q}}\,(2-\mu)}{\mu}r^{-\mu}r'^{-1-\mu}\big(r'^\mu - r^\mu\big) + \cdots\bigg] \,.
\end{align}
We also note that opposing $\pm k$ amplitudes are related by complex conjugation, e.g., $\mathcal{A}_{m \to m-1}=(\mathcal{A}_{m \to m+1})^*$.

\subsection{Axio-Dilaton} \label{subapp:axiodilaton}

The axio-dilaton two-point functions discussed in Section \ref{subsec:axiodilaton2pt} are derived with the same methods as the scalar/pseudo-scalar two-point function, though a few more considerations are in order due to the fact that our system of equations is governed by the matrix expression \eqref{LGmateq}:
\begin{align}
	e^{-h}\mathcal{L} \, G(x,x') = \delta^{(2)}_g(x-x') \qquad\text{with}\qquad \mathcal{L} = \begin{pmatrix}
		\mathcal{L}_{\tau\tau} & \mathcal{L}_{\tau\bar{\tau}} \\
		\mathcal{L}_{\bar{\tau}\tau} & \mathcal{L}_{\bar{\tau}\bar{\tau}}
	\end{pmatrix} \,, \qquad
    G = \begin{pmatrix}
		G_{\tau\tau} & G_{\tau\bar{\tau}} \\
		G_{\bar{\tau}\tau} & G_{\bar{\tau}\bar{\tau}}
	\end{pmatrix} \,.
\end{align}
Considering the explicit components
\begin{align} \label{Lcomps}
	&\mathcal{L}_{\tau\tau} = \big(\mathcal{L}_{\bar{\tau}\bar{\tau}}\big)^* = \Delta + 4\bigg(\partial_i f - \frac{\partial_i\tau}{\tau - \bar{\tau}}\bigg)\partial_i - e^{h-f}p^2 + 2\frac{\partial_i\tau\partial_i\tau}{(\tau-\bar{\tau})^2} \,, \\
    &\mathcal{L}_{\tau\bar{\tau}} = \big(\mathcal{L}_{\bar{\tau}\tau}\big)^*  = -2\frac{\partial_i\tau\partial_i\tau}{(\tau-\bar{\tau})^2} \,,
\end{align}
our first step, as in the previous section, is to redefine our operators in terms of a similarity transformation:
\begin{align} \label{LGS}
    \mathcal{L}(x) = S^{-1}(x)\widetilde{\mathcal{L}}S(x) \,, \qquad\qquad G(x,x') = S^{-1}(x)\widetilde{G}(x,x')S(x') \,,
\end{align}
with $S=\text{diag}(e^\Phi,e^{\bar{\Phi}})$, where the particular function that satisfies
\begin{align} \label{Phirels}
    \partial_i\Phi = 2\bigg(\partial_i f - \frac{\partial_i\tau}{\tau-\bar{\tau}}\bigg)
\end{align}
removes the mixed-derivative terms in our new system of equations, $\widetilde{\mathcal{L}}\widetilde{G}=\delta$.

The diagonalized derivative operator has the explicit components
\begin{align} \label{tLcomps}
	\widetilde{\mathcal{L}}_{\tau\tau} = \big(\widetilde{\mathcal{L}}_{\bar{\tau}\bar{\tau}}\big)^*
    = &\,\,\Delta - p^2\Big(r^{-2\mu} - \big(4\sqrt{q\bar{q}}(2-\mu) - 2qe^{-i\theta} - 2\bar{q}e^{i\theta}\big)r^{-1-2\mu}\Big) \nonumber\\[0.2em]
    &\,+ \sqrt{2(2-\mu)}\big(\sqrt{q}e^{-i\theta/2} + \sqrt{\bar{q}}e^{i\theta/2}\big)r^{-5/2} \nonumber\\[0.3em]
    &\,- \big(q(1+2ic_2)e^{-i\theta} + \bar{q}(1-2i\overline{c}_2)e^{i\theta}\big)r^{-3} + \cdots \,, \\[0.4em]
    \widetilde{\mathcal{L}}_{\bar{\tau}\tau} = \big(\widetilde{\mathcal{L}}_{\tau\bar{\tau}}\big)^* = &\,\,4(2-\mu)\sqrt{q\bar{q}}\,r^{-3} - 2\sqrt{2(2-\mu)}\big((4\mu-8-ic_2)q\sqrt{\bar{q}}e^{-i\theta/2} \nonumber\\[0.3em]
    &+ (4\mu-8+i\overline{c}_2)\bar{q}\sqrt{q}e^{i\theta/2}\big)r^{-7/2} \cdots \,,
\end{align}
which must now be split into $\widetilde{\mathcal{L}}_{ij,0}$ and $\delta\widetilde{\mathcal{L}}_{ij}$ in order to perform the Born approximation $(\widetilde{\mathcal{L}}_0+\delta\widetilde{\mathcal{L}})(\widetilde{G}_0+\widetilde{G}_1+\cdots)=\delta$. Writing
\begin{align} \label{G0Defs}
    \widetilde{G}_{\tau\tau,0}(x,x') = \frac{1}{2\pi}\sum_{m} e^{im(\theta - \theta')}g^{(\tau\tau)}_m(r,r') \,, \!\!\qquad  \widetilde{G}_{\bar{\tau}\bar{\tau},0}(x,x') = \frac{1}{2\pi}\sum_{m} e^{im(\theta - \theta')}g^{\bar{\tau}\bar{\tau}}_m(r,r') \,,
\end{align}
and separating out the $-m^2/r^2$ term from each of these as we did in the scalar operator case, we find that zeroth order derivative operators are both precisely equal to the derivative operator in the previous section, i.e.,
\begin{align} \label{L0mat}
    \widetilde{\mathcal{L}}_0 = \begin{pmatrix}
		\widetilde{\mathcal{L}}_{\tau\tau,0} & 0 \\
		0 & \widetilde{\mathcal{L}}_{\bar{\tau}\bar{\tau},0}
	\end{pmatrix}
    \qquad\text{with}\qquad \widetilde{\mathcal{L}}_{\tau\tau,0} = \widetilde{\mathcal{L}}_{\bar{\tau}\bar{\tau},0} = \pdv[2]{r} + r^{-1}\pdv{r} - p^2r^{-2\mu} \,.
\end{align}
This further implies that the two zeroth order Green's functions are equivalent and match what we found in the last section:
\begin{align} \label{tG0Defs}
    \widetilde{G}_{\tau\tau,0}(x,x') &= \widetilde{G}_{\bar{\tau}\bar{\tau},0}(x,x') \nonumber\\
    &= -\frac{1}{4\pi|p|}\sum_{m}e^{im(\theta' - \theta)}\,e^{-\frac{|p|}{1-\mu}(r'^{1-\mu}-r^{1-\mu})} \nonumber\\
    &\phantom{\,\,=-\frac{1}{4\pi|p|}\sum_{m}}\times(rr')^{-\frac{1-\mu}{2}}\bigg(1 - \frac{1-\mu}{8|p|}\big(r'^{-1+\mu} - r^{-1+\mu}\big) + \cdots\bigg) \,.
\end{align}

The perturbation to our derivative operator may be parameterized as
\begin{align}
    \delta\widetilde{\mathcal{L}} = \begin{pmatrix}
		U & V \\
		(V)^* & (U)^*
	\end{pmatrix} \qquad\text{with}\qquad U = \frac{1}{2\pi}\sum_{k} e^{ik\theta}u_k(r) \,, \qquad V = \frac{1}{2\pi}\sum_{k} e^{ik\theta}v_k(r) \,,
\end{align}
where the first few elements are given explicitly by
\begin{align} \label{uvkDefs}
	&u_0(r) = -2\pi\big(m^2r^{-2} + 4(2-\mu)p^2\sqrt{q\bar{q}}\,r^{-1-2\mu}\big) + \cdots \,, \\[0.4em]
    &u_{-1/2}(r) = \big(u_{1/2}(r)\big)^* = 2\pi\sqrt{2(2-\mu)}\sqrt{q}r^{-5/2} + \cdots \,, \\[0.4em]
    &u_{-1}(r) = \big(u_{1}(r)\big)^* = 2\pi q\big(p^2r^{-1-2\mu} - (1+2ic_2)r^{-3}\big)  + \cdots\,, \\[0.8em]
    &v_0(r) = 8\pi(2-\mu)\sqrt{q\bar{q}}\,r^{-3} + \cdots \,, \\[0.4em]
    &v_{-1/2}(r) = \big(v_{1/2}(r)\big)^* = 4\pi\sqrt{2(2-\mu)}(4\mu-8-ic_2)q\sqrt{\bar{q}}\,r^{-7/2} + \cdots \,, \\[0.4em]
    &v_{-1}(r) = \big(v_{1}(r)\big)^* = 2\pi q^{3/2}\sqrt{\bar{q}}\big(4c_2^2 + 32ic_2(2-\mu) - 16\mu(4\mu - 5) - 231\big)r^{-4} + \cdots \,.
\end{align}
Putting together all of the above in full matrix form, our originally complicated system of equations reduces to
\begin{align}
    \begin{pmatrix}
		\widetilde{\mathcal{L}}_{\tau\tau,0} + U & V \\
		\bar{V} & \widetilde{\mathcal{L}}_{\tau\tau,0} + \bar{U}
	\end{pmatrix}
    \begin{pmatrix}
		\widetilde{G}_{\tau\tau,0} + \widetilde{G}_{\tau\tau,1} & \widetilde{G}_{\tau\bar{\tau},1} \\
		\widetilde{G}_{\bar{\tau}\tau,1} & \widetilde{G}_{\tau\tau,0} + \widetilde{G}_{\bar{\tau}\bar{\tau},1}
	\end{pmatrix}
    =\frac{\delta(x-x')}{r}\begin{pmatrix}
		1 & 0 \\
		0 & 1
	\end{pmatrix} \,.
\end{align}
Following a treatment similar to the previous section, this system is fully determined by just three equations
\begin{align}
    &\widetilde{\mathcal{L}}_{\tau\tau,0}\widetilde{G}_{\tau\tau,0}(x,x') = \frac{\delta(x-x')}{r} \,, \\[.6em]
    &\widetilde{G}_{\tau\tau,1}(x,x') = \big(\widetilde{G}_{\bar{\tau}\bar{\tau},1}(x,x')\big)^* \nonumber\\
    &\phantom{\widetilde{G}_{\tau\tau,1}(x,x')}= \frac{1}{(2\pi)^2}\int_{r_0}^\infty \dd{r_s}r_s\sum_{k,m}g(r,r_s)u_k(r_s)g(r_s,r')e^{i(m\theta - (m-k)\theta')} \,, \\
    &\widetilde{G}_{\tau\bar{\tau},1}(x,x') = \big(\widetilde{G}_{\bar{\tau}\tau,1}(x,x')\big)^* \nonumber\\
    &\phantom{\widetilde{G}_{\tau\tau,1}(x,x')}= \frac{1}{(2\pi)^2}\int_{r_0}^\infty \dd{r_s}r_s\sum_{k,m}g(r,r_s)v_k(r_s)g(r_s,r')e^{i(m\theta - (m-k)\theta')}\,.
\end{align}

With the solution to the first of these already in hand, \eqref{tG0Defs}, solutions to the remaining two equations follow by simply computing the integrals and collecting terms. All that remains is to assemble the complete Green's functions and convert back to the original un-tilded basis through the relations \eqref{LGS}, where the explicit expansion of the transformation function is given by
\begin{align}
    e^{\Phi(x)} =&\,\,1 - 2\sqrt{2(2-\mu)}\big(\sqrt{q}e^{-i\theta/2} + \sqrt{\bar{q}}e^{i\theta/2}\big)r^{-1/2} \nonumber\\
    &- \frac12\big((8\mu - 2ic_2-17)qe^{-i\theta} - (8\mu + 2i\overline{c}_2-17)\bar{q}e^{i\theta}\big)r^{-1} + \cdots \,,
\end{align}
per our known background solutions and the relations \eqref{Phirels}. Here we encounter a crucial difference between the present calculation and the free scalar case -- the combination of periodic and anti-periodic components in the potential of \eqref{Lcomps} induces a twisting that manifests as a sum over $k,m\in\mathbb{Z}/2$, as discussed in the main text. We see this appear explicitly in the complete Green's functions which take the form
\begin{align}
	G_{ij}(y,y') = \sum_{k,m}\mathcal{A}^{(ij)}_{m \to m-k}(r,r')e^{ik\theta'}e^{im(\theta'-\theta)} \,.
\end{align}
The first few diagonal amplitudes, shown here to higher order than in \eqref{tt0trans}--\eqref{tt1trans}, are given by
{\allowdisplaybreaks
\begin{align}
	\mathcal{A}^{(\tau\tau)}_{m \to m}(r,r') = &-\frac{1}{4\pi}\,e^{-\frac{|p|}{1-\mu}(r'^{1-\mu}-r^{1-\mu})}(rr')^{-\frac{1-\mu}{2}} \nonumber\\
    & \qquad \times\bigg[\frac{1}{|p|} - \frac{m^2}{4\pi p^2(1-\mu)}(rr')^{-1}\big(r'^{\mu} - r^{\mu}\big) \nonumber\\
    & \qquad\qquad + \frac{\sqrt{q\bar{q}}\,(2-\mu)}{\pi\mu}(rr')^{-\mu}\big(r'^\mu - r^\mu\big) + \cdots\bigg] \,, \\
	\mathcal{A}^{(\tau\tau)}_{m \to m+\frac12}(r,r') = &\,\,\frac{\sqrt{q\,(2-\mu)}}{\sqrt{2}\pi}e^{-\frac{|p|}{1-\mu}\big(r'^{1-\mu}-r^{1-\mu}\big)}(rr')^{-\frac{1-\mu}{2}}r'^{-1/2} \nonumber\\
    & \qquad \times\bigg[\frac{1}{|p|} - \frac{m^2}{4\pi p^2(1-\mu)}\big(r'^{-1+\mu} - r^{-1+\mu}\big) \nonumber\\
    & \qquad\qquad - \frac{\sqrt{q\bar{q}}(2\mu-3)}{2\pi\mu}(rr')^{-\mu}\big(r'^\mu - r^\mu\big) + \cdots \bigg] \,, \\
	\mathcal{A}^{(\tau\tau)}_{m \to m+1}(r,r') = &-\frac{q}{8\pi^2}\,e^{-\frac{|p|}{1-\mu}(r'^{1-\mu}-r^{1-\mu})}(rr')^{-\frac{1-\mu}{2}} \nonumber\\
    & \qquad \times\bigg[\frac{1}{\mu}\big(r'^{-\mu} - r^{-\mu}\big) - \frac{m^2(8\mu - 2ic_2-17)}{4p^2(1-\mu)}r'^{-1}\big(r'^{-1+\mu} - r^{-1+\mu}\big) \nonumber\\
    & \qquad\qquad - \frac{\sqrt{q\bar{q}}\,(2-\mu)}{\mu}(rr')^{-1-\mu}\big(r'^\mu-r^\mu\big)\big(16r' + (8\mu - 2ic_2-17)r\big) + \cdots\bigg] \,.
\end{align} }
Next-to-leading expressions of the first couple of off-diagonal amplitudes shown in \eqref{ttb0trans}--\eqref{ttb12trans} take the forms
\begin{align}
	\mathcal{A}^{(\tau\bar{\tau})}_{m \to m}(r,r') = &\,\,\frac{\sqrt{q\bar{q}}}{4\pi^2p^2}e^{-\frac{|p|}{1-\mu}(r'^{1-\mu}-r^{1-\mu})}(rr')^{-\frac{1-\mu}{2}} \nonumber\\
    &\qquad \!\times\bigg[\big(r'^{-2+\mu} - r^{-2+\mu}\big) + \frac{1-\mu}{8|p|}(rr')^{-2+\mu}(r + r') \nonumber\\
    &\qquad\qquad - \sqrt{q\bar{q}}(2-\mu)r'^{-2+\mu}r + \cdots\bigg] \,, \\
	\mathcal{A}^{(\tau\bar{\tau})}_{m \to m+\frac12}(r,r') = &\,\,\frac{q\sqrt{\bar{q}\,(2-\mu)}}{\sqrt{2}\pi^2p^2}e^{-\frac{|p|}{1-\mu}\big(r'^{1-\mu}-r^{1-\mu}\big)}(rr')^{-\frac{1-\mu}{2}} \nonumber\\
    &\qquad \!\times\bigg[r'^{-1/2}r^{-2+\mu} - \frac{1-\mu}{8|p|}r'^{-\frac{5-2\mu}{2}}r^{-2+\mu}(r + r') \nonumber\\
    &\qquad\qquad + \frac{8\sqrt{q\bar{q}}(2-\mu)(ic_2 + 4\mu - 8)}{2\mu-5}r^{-\frac{7+2\mu}{2}} + \cdots\bigg] \,, \\
    \mathcal{A}^{(\tau\bar{\tau})}_{m \to m+1}(r,r') = &-\frac{q^{3/2}\sqrt{\bar{q}}}{\pi^2p^2}e^{-\frac{|p|}{1-\mu}\big(r'^{1-\mu}-r^{1-\mu}\big)}(rr')^{-\frac{1-\mu}{2}} \nonumber\\
    &\qquad \!\times\bigg[\frac{(2-\mu)(ic_2 + 4\mu - 8)}{2\mu - 5}r'^{-1/2}r^{-\frac{5-2\mu}{2}} + \frac{(1-\mu)^{3/2}}{8\sqrt{2}|p|}r'^{-\frac{5-2\mu}{2}}r^{-2+\mu}(r'+ r)\nonumber\\
    &\qquad\qquad - \frac{4\sqrt{q\bar{q}}(2-\mu)^{3/2}(ic_2 + 4\mu - 8)}{2\mu - 5}r^{-\frac{7-2\mu}{2}} + \cdots\bigg] \,,
\end{align}
which are suppressed relative to the diagonal amplitudes by a factor of $r^{-2+\mu}$ or more.

\section{An $\mathrm{AdS}_9$ Limit}

Here, for the sake of general interest, we note that there exists a limit of our 10D metric solutions presented in the main text that precisely reproduces the metric of $\text{AdS}_9$. This metric can be expressed in Poincar\'{e} coordinates as
\begin{align} \label{AdS9}
    \dd{s}^2 = \frac{L^2}{w^2}\big(\eta_{ab}\dd{x}^a\dd{x}^b + \dd{w}^2\big) \,,
\end{align}
where $w$ is a bulk radial coordinate and $L$ is the AdS radius. The metric ansatz expressed in polar coordinates takes the form
\begin{align} \label{polaransatz}
    \dd{s}^2 = e^{f(r,\theta)}\eta_{ab}\dd{x}^a\dd{x}^b + e^{h(r,\theta)}\big(\dd{r}^2 + r^2\dd{\theta}^2\big) \,.
\end{align}
If we limit ourselves to a particular ray in the $\theta$-direction by setting a constant $\theta=\theta_c$, it becomes readily apparent that a coordinate transformation $w(r)$ that satisfies
\begin{align} \label{wtrans}
    e^{f(r,\theta_c)} = \frac{L^2}{w(r)^2} \,, \qquad\qquad\qquad e^{h(r,\theta_c)}\dd{r}^2 = \frac{L^2}{w(r)^2}\dd{w}^2 \,,
\end{align}
directly equates \eqref{AdS9} to \eqref{polaransatz} along the ray $\theta=\theta_c$. Given our previously calculated solution for $f$, \eqref{fSol}, this explicit coordinate relationship is:
\begin{align}
    w(r) = \frac{Lr^{1/8}}{(r + \omega_c)^{1/8}} \,, \qquad\text{where}\qquad \omega_c = 2q_x\cos\theta_c \,.
\end{align}
In terms of our transverse radial coordinate $r$, the conformal boundary is approached as $r\to0$, while the AdS horizon is found at the particular point $r=-\omega_c$.

The transformation defined by \eqref{wtrans} also implies an exact solution for $h$ that, along with its large-$r$ expansion, is given by
\begin{align}
    h(r,\theta_c) &= 2\log(\frac{L\omega_c}{8r(r+\omega_c)}) = 2\log(\frac{L\omega_c}{8}) - 4\log r - \frac{2\omega_c}{r} + \cdots \,.
\end{align}
Interestingly, this form for $h$ can be made to match our previously found solution exactly, for a given set of parameters. To see equivalency, we must add an arbitrary constant $b_0$ to our $h$ series \eqref{hSol}, which, given the leading logarithmic term in this solution, does not affect the asymptotic behavior of any of our previous results. With this, we can simply match terms in the large-$r$ expansion of the $h$ solution above to find that they are equivalent after setting
\begin{align}
    \mu = 2 \,, \qquad\qquad\qquad b_0 = 2\log(\frac{L\omega_c}{8}) \,,
\end{align}
and after fixing the arbitrary constant $c_2$ so that it satisfies
\begin{align}
    \frac{1}{4r}\Big(8\omega_c - \big(2c_2{}^2 + 2\overline{c}_2^2 + 7\big) - 2i(c_2 + \overline{c}_2)(c_2 - \overline{c}_2)\Big) = 0 \,.
\end{align}
We note that it appears that this equivalency may be established at all higher orders in the expansion as well, due to the extra $c_i$ terms that appear at each order and the fact that the value of $\mu=2$ kills many terms in the expansion.

This value of the tension is notable because it saturates the Elementary Constituents Conjecture \cite{Nevoa:2025xiq} bound and implies over-closure of the background when our transverse space is interpolated to a closed surface. This potential issue aside, it is interesting to highlight that it becomes possible to make the statement that our metric can be made to look like AdS in the limit $\mu\to2$ for any fixed $\theta=\theta_c$. This is of course not the statement that our metric becomes AdS at this value globally, just that there exists a family of solutions where our spacetime becomes AdS on any given $\theta$-ray.

Though this parallel between our solution and $\text{AdS}$ backgrounds suggests the possibility of drawing comparisons to well-known results in holography, e.g., CFT correlators, there exists a significant obstruction in the fact that our solution cannot access the conformal boundary at $r=0$. While it would be interesting to generalize the present model and attempt to draw parallels to AdS/CFT, such considerations are beyond the scope of this work.

\newpage

\bibliographystyle{utphys}
\bibliography{R7metric}

\end{document}